\documentclass[aps,prd,twocolumn,showpacs,nofootinbib,amsmath,amssymb,amsfonts]{revtex4}
\usepackage{graphicx,epsfig,rotate,psfig}   
\usepackage{dcolumn}      
\usepackage{bm}           

\newcommand{\scalefac}{R}
\newcommand{\baryon}{{\rm b}}
\newcommand{\xout}{{\rm out}}
\newcommand{\xin}{{\rm in}}
\newcommand{\obh}{$\Omega_{\baryon}h^2$}

\newcommand{\deu}{${\rm D}$}

\newcommand{\qua}{$^4{\rm He}$}

\newcommand{\sep}{$^{7}{\rm Li}$}
\newcommand{\cav}{\mathrm{cav}}
\newcommand{\eff}{\mathrm{eff}}
\newcommand{\ppn}{\mathrm{PPN}}

\newcommand{\dd}{\mathrm{d}}

\newcommand\beq{\begin{equation}}
\newcommand\eeq{\end{equation}}
\def\iso#1#2{\mbox{${}^{#2}{\rm #1}$}}

\def\he#1{\iso{He}{#1}}
\def\li#1{\iso{Li}{#1}}

\begin{document}

\leftline{UMN--TH--2424/05, FTPI--MINN--05/52}

\title{Big bang nucleosynthesis constraints on scalar-tensor theories of gravity}

\author{Alain Coc}
 \email{coc@csnsm.in2p3.fr}
 \affiliation{Centre de Spectrom\'etrie Nucl\'eaire et de
Spectrom\'etrie de Masse, IN2P3/CNRS/UPS,
B\^at. 104, 91405 0rsay Campus (France)}

\author{Keith A. Olive}
 \email{olive@physics.unm.edu}
 \affiliation{William I. Fine Theoretical Physics Institute,
              University of Minnesota, Minneapolis, MN 55455 (USA)}

\author{Jean-Philippe Uzan}
 \email{uzan@iap.fr}
 \affiliation{Institut d'Astrophysique de Paris,
              UMR-7095 du CNRS, Universit\'e Pierre et Marie
              Curie,
              98 bis bd Arago, 75014 Paris (France)}

\author{Elisabeth Vangioni}
 \email{vangioni@iap.fr}
 \affiliation{Institut d'Astrophysique de Paris,
              UMR-7095 du CNRS, Universit\'e Pierre et Marie
              Curie,
              98 bis bd Arago, 75014 Paris (France)}
\date{\today}
\begin{abstract}
We investigate BBN in scalar-tensor theories of gravity with arbitrary
matter couplings and self-interaction potentials. We first consider the
case of a massless dilaton with a quadratic coupling to matter. We
perform a full numerical integration of the evolution of the
scalar field and compute the resulting light element abundances.
We demonstrate in detail the importance of particle mass
thresholds on the evolution of the scalar field in a radiation
dominated universe. We also consider the simplest extension of
this model including a cosmological constant in either the Jordan
or Einstein frame.
\end{abstract}

\pacs{PACS}
\maketitle

\section{Introduction}\label{sec1}

The concordance model of cosmology calls for the introduction of a
cosmological constant or a dark energy sector. Various
candidates have been proposed~\cite{darkrevue}, among which the
possibility that gravity is not described by general relativity on
large cosmological scales. It is of interest therefore, to test
our theory of gravity in a cosmological context. This can be
achieved in two complementary ways, either by designing model
independent tests (see e.g. Refs.~\cite{uam,ctes2,bct} for a
discussion of the various possible tests) or by considering a
class of well motivated theories and use all available data to
determine how close to general relativity we must be.

Among all extensions of general relativity, scalar-tensor theories
are probably the simplest in the sense that they consider only the
introduction of one~\cite{stgen} (or many~\cite{def}) scalar
field(s) universally coupled to matter. These theories involve two free
functions describing the coupling of the scalar field to matter
and its self-interaction potential.
They respect 
local Lorentz invariance and the universality of free fall of
laboratory-size bodies. They are motivated by high-energy theories
trying to unify gravity with other interactions which generically
involve a scalar field in the gravitational sector. In particular,
in superstring theories~\cite{polchy} the supermultiplet of the
10-dimensional graviton contains a scalar field, the dilaton, and
other scalar fields, moduli, appear during Kaluza-Klein
dimensional reduction of higher dimensional theories to our usual
four dimensional spacetime.

In cosmology, two main properties make these theories appealing.
First, an attraction mechanism toward general
relativity~\cite{dn1,dpol} was exhibited. This implies that even
if the tests of general relativity in the Solar system set strong
constraints on these theories, they may differ significantly from
general relativity at high redshift. Second, it was shown that the
general mechanism of quintessence was
conserved~\cite{jpu99,autres} if the quintessence field was
non-minimally coupled and that the attraction mechanism toward
general relativity still held with runaway
potentials~\cite{bp,runaway1,runaway2}. These extended
quintessence models are the simplest theories in which there is a
long range modification of gravity, since the quintessence field
is light, and they allow for a very interesting
phenomenology~\cite{msu5}.

Cosmological data give access to various aspects of these models.
The cosmic microwave background (CMB) tests the theory in the linear
regime~\cite{ru02,cmb1,cmb2,cmb3,cmb4} while weak lensing opens a
complementary window on the non-linear regime~\cite{carlo,acqua}.
Solar system experiments give information on the theory today and
big-bang nucleosynthesis (BBN) allows us to constrain the attraction
mechanism toward general relativity~\cite{dn1} at very high redshift.\\

BBN is one of the most sensitive available probes of the very
early Universe and of physics beyond the standard model. Its
success rests on the concordance between the observational
determinations of the light element abundances of D, \he3, \he4,
and \li7, and their theoretically predicted abundances \cite{bbn,
bbnb}. Furthermore, measurements of the CMB anisotropies by WMAP
\cite{wmap} have led to precision determinations of the baryon
density or equivalently the baryon-to-photon ratio, $\eta$. As
$\eta$ is the sole parameter of the standard model of BBN,
 it is possible to make very accurate predictions \cite{cfo3,coc,pdes,cuoco,cyburt}
 and hence further
 constrain physics beyond the standard model \cite{cfos}.

In particular, the \he4 abundance is often used as a sensitive
probe of new physics.  This is due to the fact that nearly all
available neutrons at the time of BBN end up in \he4 and the
neutron-to-proton ratio is very sensitive to the competition between
the weak interaction rate and the expansion rate.  Of interest to us here
is the effect of modifications to gravity which will directly affect the expansion rate
of the Universe through a modified Friedmann equation.

The WMAP best fit assuming a varying spectral index is
$\Omega_\baryon h^2 = 0.0224 \pm 0.0009$ and is equivalent to
$\eta_{\rm 10,{\rm CMB}} = 6.14 \pm 0.25$, where $\eta_{10} =
10^{10} \eta$. Using the WMAP data to fix the baryon density,
the light element abundances \cite{cfo3,coc,pdes,cuoco,cyburt} can be
quite accurately predicted. Some BBN results are displayed in Table
1.

\begin{table*}
\caption{BBN results for the light element abundances assuming
the WMAP-inferred baryon density.}
\begin{center}
\begin{tabular}{l|c|c|c|c}
 \hline\hline
 Source& $Y_p$ & ${\rm D}/{\rm H}$&$^3{\rm He}/{\rm H}$ &
       $^{7}{\rm Li}/{\rm H}$\\ &&$\times10^{-5}$&$\times10^{-5}$&$\times10^{-10}$\\
 \hline\hline
 Coc {\it et al.} (2004)& 0.2479$\pm0.0004$&$2.60\pm0.17$&$1.04\pm0.04$&$4.15\pm0.46$\\
\hline
 Cyburt {\it et al.} (2003) & $0.2484^{+0.0004}_{-0.0005}$ & $2.74^{+0.26}_{-0.16}$ & $0.93^{+0.1}_{-0.67}$ & $3.76^{+1.03}_{-0.38}$\\
\hline\hline
\end{tabular}
\end{center}
\label{t:yields}
\end{table*}

The effect of scalar-tensor theories of gravity on the production
of light elements has been investigated extensively (see e.g.
Ref.~\cite{ctes} for a review). As a first step, it is useful to
consider only the speed up factor, $\xi=H/H_{\rm GR}$, that
arises from the modification of the value of the gravitational
constant during BBN~\cite{speedup1,cfos}. Other approaches
considered the
full dynamics of the problem but restricted themselves to the particular class of
Jordan-Fierz-Brans-Dicke theory~\cite{bbnJFBD}, of a massless
dilaton with a quadratic coupling~\cite{dp,bbn_quad} or to a general
massless dilaton~\cite{bbnst1}. It should be noted that a combined
analysis of BBN and CMB data was investigated in Ref.~\cite{copi}
and Ref.~\cite{bbnst2}. The former considered $G$ constant during
BBN while the latter focused on a non-minimally quadratic coupling
and a runaway potential. We stress that the dynamics of the
field can modify CMB results so that one needs to be careful
while inferring $\Omega_\baryon$ from WMAP.

The goal of this article is to implement scalar-tensor theories in
an up-to-date BBN code. This will complement our existing set of
tools which allows us to confront scalar-tensor theories with
observations of  type Ia supernovae  and CMB
anisotropies~\cite{ru02} as well as weak lensing~\cite{carlo}. In
particular, the predictions to be compared with observations can
be computed in the same framework for any self-interaction
potential and matter-coupling function.

We first recall, in \S~\ref{sec_ST_theory}, the equations
describing the theory to be implemented in our BBN code and we
also discuss local constraints. As a check of our code, we
consider, in \S~ \ref{sec_modelDP}, the case of a massless dilaton
with quadratic coupling~\cite{dp}. In particular, we perform a
full numerical integration up to the present that can be compared
with the analytical results of Ref.~\cite{dp}. We update the
constraints on this model by taking into account the latest BBN
data discussed above. We reaffirm that only helium-4 is sensitive
to the modification of gravity considered here. In
\S~\ref{sec_model2}, we will consider the simplest extension of
this model by introducing a cosmological constant. Such a constant
can be introduced as a constant potential either in the Einstein
frame, hence keeping the dilaton massless, or in the Jordan frame,
hence generalizing the constant energy density component. Both
cases are considered and we conclude in Section~\ref{sec_concl}.
Applications to various cases of cosmological interest will be
presented in a follow-up article.

\section{Implementing scalar-tensor theories of gravity in a BBN
code}\label{sec_ST_theory}

\subsection{Scalar-tensor theories in brief}

In  scalar-tensor theories of gravity,
gravity is mediated not only by a spin-2
graviton but also by a spin-0 scalar field that couples
universally to matter fields. In the Jordan frame, the action
of the theory takes the form
\begin{eqnarray}\label{actionJF}
  S &=&\int \frac{\dd^4 x }{16\pi G_*}\sqrt{-g}
     \left[F(\varphi)R-g^{\mu\nu}Z(\varphi)\varphi_{,\mu}\varphi_{,\nu}
        - 2U(\varphi)\right]\nonumber\\
        &&   \qquad+ S_m[g_{\mu\nu};\psi]
\end{eqnarray}
where $G_*$ is the bare gravitational constant from which we
define $\kappa_*=8\pi G_*$. This action involves three arbitrary
functions ($F$, $Z$ and $U$) but  only two are physical
since there is still the possibility to redefine the
scalar field. $F$ needs to be positive to ensure that the graviton
carries positive energy. $S_m$ is the action of the matter fields
that are coupled minimally to the metric $g_{\mu\nu}$ with
signature $(-,+,+,+)$.

The action~(\ref{actionJF}) can be rewritten in the Einstein frame
by performing the conformal transformation
\begin{equation}\label{jf_to_ef}
 g_{\mu\nu}^* = F(\varphi)g_{\mu\nu}
\end{equation}
as
\begin{eqnarray}\label{actionEF}
 S &=& \int \frac{\dd^4x}{16\pi G_*}\sqrt{-g_*}\left[ R_*
        -2g_*^{\mu\nu} \partial_\mu\varphi_*\partial_\nu\varphi_*
        - 4V(\varphi_*)\right]\nonumber\\
  &&\qquad + S_m[A^2(\varphi_*)g^*_{\mu\nu};\psi].
\end{eqnarray}
The field $\varphi_*$ and the two functions $A(\varphi_*)$ and
$V(\varphi_*)$ are defined by
\begin{eqnarray}
 \left(\frac{\dd\varphi_*}{\dd\varphi}\right)^2
              &=& \frac{3}{4}\left[\frac{\dd\ln F(\varphi)}{\dd\varphi}\right]^2
                  +\frac{Z(\varphi)}{2F(\varphi)}\label{jf_to_ef1}\\
 A(\varphi_*) &=& F^{-1/2}(\varphi)\label{jf_to_ef2}\\
 2V(\varphi_*)&=& U(\varphi) F^{-2}(\varphi)\label{jf_to_ef3}.
\end{eqnarray}
We will denote any Einstein frame quantities by a star (*), e.g.
$R_*$ is the Ricci scalar of the metric $g^*_{\mu\nu}$. The
strength of the coupling of the scalar field to the matter fields
is characterized by
\begin{equation}\label{eqalpha}
 \alpha(\varphi_*)\equiv \frac{\dd\ln A}{\dd\varphi_*}
\end{equation}
and we also define
\begin{equation}\label{eqbeta}
 \beta(\varphi_*)\equiv \frac{\dd\alpha}{\dd\varphi_*}.
\end{equation}

It is useful to study both the Einstein and Jordan frames. In the
Jordan frame, matter is universally coupled to the metric. The
Jordan metric defines the length and time as measured by
laboratory apparatus so that all observations (time, redshift,...)
have their standard interpretation in this frame. However, to
discuss the theory it is often better to use the Einstein frame in
which the kinetic terms have been diagonalized so that the spin-2
and spin-0 degrees of freedom of the theory are perturbations of
$g^*_{\mu\nu}$ and $\varphi_*$ respectively. The physical
properties of both frames are of course identical. For example,
when we refer to the time variation of  the gravitational constant
(in the Jordan frame), we have assumed fixed particle masses. In
contrast, in the Einstein frame, we would infer a fixed
gravitational constant and varying masses.  In both frames, the
quantity $Gm^2$ (which is physically measureable) varies in the
same way.

\subsection{Friedmann equations}

\subsubsection{Equations in Jordan frame}

We consider a Friedmann-Lema\^{\i}tre universe with metric in the
Jordan frame
\begin{equation}
 \dd s^2 = -\dd t^2 + \scalefac^2(t)\gamma_{ij}\dd x^i \dd x^j
\end{equation}
where $\gamma_{ij}$ is the spatial me\-tric and $\scalefac$ the
scale factor. The matter fields are described by a collection of
perfect fluids of energy density, $\rho$ and pressure $P$. It
follows that the Friedmann equations in Jordan frame take the form
\begin{eqnarray}
 3F\left(H^2 + \frac{K}{\scalefac^2}\right) &=& 8\pi G_*\rho
  +\frac{1}{2}Z\dot\varphi^2-3H\dot F + U \label{einsteinJF1}\\
 -2F\left(\dot H - \frac{K}{\scalefac^2}\right) &=& 8\pi G_*(\rho+P)
  + Z\dot\varphi^2\nonumber\\
  &&\qquad +\ddot F - H\dot F\label{einsteinJF2}
\end{eqnarray}
where a dot refers to a derivative with respect to the cosmic time
$t$ and $H\equiv\dd\ln\scalefac/\dd t$. The Klein-Gordon and
conservation equations are given by
\begin{eqnarray}
 && Z(\ddot\varphi+3H\dot\varphi)=3F_{\varphi}\left(\dot H + 2H^2
 +\frac{K}{\scalefac^2}\right)\nonumber\\
 &&\qquad\qquad - \frac{1}{2}Z_{\varphi}\dot\varphi^2 -
 U_{\varphi}\label{kleinJF1}\\
 &&\dot\rho+3H(\rho+P) = 0.
\end{eqnarray}
If we define the density parameters today by
\begin{equation}
 \Omega_0 \equiv \frac{8\pi G_*\rho_0}{3H_0^2 F_0},
\end{equation}
the evolution of the energy density of a fluid with constant
equation of state $w=P/\rho$ takes the usual form
\begin{equation}
 \rho = \frac{3H_0^2 F_0\Omega_0}{8\pi G_*}(1+z)^{3(1+w)}
\end{equation}
where $z$ is the redshift defined by $1+z=\scalefac_0/\scalefac$.

\subsubsection{Equations in Einstein frame}

The scale factor and cosmic time in Einstein frame are related to
the ones in Jordan frame by
\begin{equation}\label{JF2EF}
 \scalefac = A(\varphi_*) \scalefac_*,\qquad
 \dd t = A(\varphi_*)\dd t_*
\end{equation}
so that the redshifts are related by
\begin{equation}
 1+z=\frac{A_0}{A}(1+z_*).
\end{equation}

The Friedmann equations in this frame take the form
\begin{eqnarray}
 && 3\left(H_*^2 + \frac{K}{\scalefac_*^2}\right) =
   8\pi G_*\rho_* + \psi_*^2 + 2V(\varphi_*)\label{einsteinEF1}\\
 && - \frac{3}{\scalefac_*^2}\frac{\dd^2 \scalefac_*}{\dd t_*^2} = 4\pi G_* (\rho_* + 3P_*)
   + 2 \psi_*^2 - 2V(\varphi_*)\label{einsteinEF2}
\end{eqnarray}
where we have introduced $H_*=\dd \ln \scalefac_*/\dd t_*$ and
\begin{equation}\label{defpsistar}
 \psi_* = \dd\varphi_*/\dd t_*.
\end{equation}
These equations take the same form as the standard Friedmann
equations for a universe containing a perfect fluid and a scalar
field. The Klein-Gordon equation takes the form
\begin{equation}\label{kgEF}
 \frac{\dd\psi_*}{\dd t_*} + 3H_*\psi_* = -\frac{\dd V}{\dd\varphi_*}
  -4\pi G_*\alpha(\varphi_*)(\rho_*-3P_*)
\end{equation}
while the matter conservation equation is given by
\begin{equation}\label{evEF}
 \frac{\dd\rho_*}{\dd t_*} + 3H_*(\rho_* + P_*) =
 \alpha(\varphi_*)(\rho_*-3P_*)\psi_*.
\end{equation}
These equations differ from their standard form due to the
coupling that appears in the r.h.s. The solution of the evolution
equation~(\ref{evEF}) can be obtained from the relation between
the energy density and the pressure of a fluid in Einstein frame
and their Jordan frame counterparts
\begin{equation}
 \rho_* = A^4\rho,\qquad P_* = A^4 P
\end{equation}
which imply, in particular, that
\begin{equation}
 \rho_* = \frac{3H_0^2\Omega_0}{8\pi G_*}\left(\frac{A}{A_0}\right)^{4-3(1+w)}(1+z_*)^{3(1+w)}
\end{equation}
for a fluid with a constant equation of state.

\subsection{Constraints today}

\subsubsection{Post-newtonian constraints}

The post-Newtonian parameters (see Refs.~\cite{will,gefp}) can be
expressed in terms of the values of $\alpha$ and $\beta$ today as
\begin{equation}
 \gamma^\ppn - 1 = -\frac{2\alpha_0^2}{1+\alpha^2_0},\qquad
 \beta^\ppn - 1 =\frac{1}{2}
 \frac{\beta_0\alpha_0^2}{(1+\alpha_0^2)^2}.
\end{equation}
Solar System experiments set strong limits on these parameters.
The perihelion shift of Mercury implies~\cite{mercurybound}
\begin{equation}
 |2\gamma^\ppn - \beta^\ppn -1|<3\times10^{-3},
\end{equation}
the Lunar Laser Ranging experiment~\cite{llrbound} sets
\begin{equation}
  4\gamma^\ppn - \beta^\ppn -3 = -(0.7\pm1)\times10^{-3}.
\end{equation}
Two experiments give a bound on $\gamma^\ppn$ alone, the Very Long
Baseline Interferometer~\cite{vlbibound}
\begin{equation}
 |\gamma^\ppn -1|< 4\times10^{-4},
\end{equation}
and the measurement of the time delay variation to the Cassini
spacecraft near Solar conjunction~\cite{cassinibound}
\begin{equation}
 \gamma^\ppn -1 = (2.1\pm2.3)\times10^{-5}.
\end{equation}
These two last bounds imply $\alpha_0$ to be very small, typically
$\alpha_0^2<10^{-5}$ while $\beta_0$ can still be
large~\cite{pulsar}. Binary pulsar observations impose that
$\beta_0\gtrsim-4.5$. Note that even though $\beta_0$ is not
bounded above by experiment, we will assume that it is not very
large, typically we assume $\beta_0\lesssim100$, so that the
post-Newtonian approximation scheme makes sense.

\subsubsection{Gravitational constant}

The Friedmann equations in the Jordan frame
define an effective gravitational constant
\begin{equation}
 G_\eff = G_*/F = G_*A^2.
\end{equation}
This constant, however,  does not correspond to the gravitational constant
effectively measured in a Cavendish experiment. The
constant measured in this type of experiment is
\begin{equation}
 G_\cav = G_*A_0^2(1+\alpha_0^2)
\end{equation}
where the first term, $G_*A_0^2$, corresponds to the exchange of a
graviton while the second term, $G_*A_0^2\alpha_0^2$, is related
to the long range scalar force.

Assuming fixed particle masses, the time variation of the gravitational constant is
bounded~\cite{dickey} by
\begin{equation}
 \frac{1}{G_\cav}\frac{\dd G_\cav}{\dd t} = \sigma_0 H_0,\qquad
 |\sigma_0| < 5.86\times10^{-2}h^{-1}.
 \label{gconst0}
\end{equation}
Choosing the number of Einstein frame $e$-folds as a time variable,
\begin{equation}\label{def_p}
 p = -\ln(1+z_*),
\end{equation}
implies that
\begin{equation}\label{gconst}
 2\alpha_0\left[1+\frac{\beta_0}{1+\alpha_0^2}-\frac{\sigma_0}{2} \right]
 \left.\frac{\dd\varphi_*}{\dd p}\right|_0=\sigma_0.
\end{equation}
Note that the limit $\beta=-(1+\alpha^2)$ that corresponds to the
so-called Barker theory~\cite{barker} in which $A=\cos\varphi_*$
leads to $\sigma=0$ whatever the value of $\alpha$ and
$\varphi_*'$ so that the gravitational constant is strictly
constant even though gravity is not described by general
relativity.

\subsection{Numerical implementation}
\label{s:num}

The nuclear reaction network takes its standard form in Jordan
frame. To compute the light elements abundances during BBN, one
only needs to know the expansion rate history, $H(z)$, from deep
in the radiation era up to today. It is thus convenient to express
the Hubble parameter in the Jordan frame in terms of the one in the
Einstein frame, using Eq.~(\ref{JF2EF}), as
\begin{equation}\label{HHstar}
 AH = \left[H_* + \alpha(\varphi_*)\psi_*\right]
\end{equation}
where $\psi_*$ is defined by Eq.~(\ref{defpsistar}). Eq.
(\ref{HHstar}) can also be expressed in the simple form
\begin{equation}
 AH = H_*\left[1 + \alpha(\varphi_*)\frac{\dd\varphi_*}{\dd
 p}\right].
\end{equation}

It follows that, in terms of the cosmic time $t$, the equations of
evolution can be recast as
\begin{eqnarray}
 \frac{\dd\varphi_*}{\dd t} &=& A^{-1}(\varphi_*) \psi_*\label{s1} \\
 \frac{\dd\psi_*}{\dd t} &=& -A^{-1}(\varphi_*)\left[3H_*\psi_* \right.\nonumber\\
 &&\!\!\!\!\!\!\!\!\!\!+\left.4\pi G_*\alpha(\varphi_*)A^4(\varphi_*)\sum_i
 (1-3w_i)\rho_i + \frac{\dd V}{\dd\varphi_*} \right]\\
 H^2_* &=&\!\! \frac{8\pi G_*}{3}A^4(\varphi_*)\sum_i\!\rho_i +
 \!\frac{1}{3}\psi_*^2\! + \!\frac{2}{3}V(\varphi_*)\! -\!
 \frac{K}{\scalefac^2_*}\\
 \rho_i &=& \rho_{i0}(1+z)^{3(1+w_i)}\\
 H &=& A^{-1}\left[H_* + \alpha(\varphi_*)\psi_*\right].\label{s5}
\end{eqnarray}

The numerical integration is performed as follows. First we choose
some initial value $\varphi_{\rm in*}$, $\psi_{\rm in*}=0$ deep in
the radiation era (typically, $z_{\rm in}= 10^{12}$ and we
integrate the system~(\ref{s1}-\ref{s5}) to $z=0$. We perform a
shooting method so that the solution reaches the value
$\Omega_{\Lambda0}$ and $G_N$ today, which fixes $G_*$ and the
energy scale of the potential. At this stage the value of
$\varphi_{*0}$ and $\alpha_0$ are known. We also keep track of
$\psi_{0*}$ to infer the time variation of the gravitational
constant, $G_\cav$, and check its compatibility with the
constraint~(\ref{gconst0}). Subsequently, we perform a second
integration of the same system including the nuclear reaction
network.

\section{Massless dilaton with quadratic coupling}\label{sec_modelDP}

The simplest model to consider consists of a massless dilaton with a
quadratic coupling to matter. That is,
\begin{equation}\label{mod1}
 V(\varphi_*) = 0, \qquad A=\hbox{e}^{a(\varphi_*)},\qquad
 a(\varphi_*) = \frac{1}{2}\beta\varphi_*^2.
\end{equation}
It follows that
\begin{equation}
 \alpha_0 = \beta\varphi_{0*},\qquad
 \beta_0  = \beta.
\end{equation}
This model has been studied in detail in the literature, both in
terms of its dynamics~\cite{dn1,dpol} and of its BBN
predictions~\cite{dp,bbn_quad}. We use it as a test model to check
our numerical scheme. In particular, the analytical behaviour of the field
during the radiation and matter eras after BBN was obtained
for a flat universe without cosmological constant in
Ref.~\cite{dp}. The numerical integration through BBN
was also matched to this solution. Since we would like
to use the same integration scheme
for any potential and coupling, we can not rely on a
particular analytic solution. It is used in this particular
case only to check the accuracy of our code.

\subsection{General study}

As long as $V=0$, the Klein-Gordon equation~(\ref{kgEF}) can be
rewritten in terms of the variable $p$ defined by
Eq.~(\ref{def_p}) as
\begin{eqnarray}\label{kgqq}
 \frac{2}{3-\varphi_*^{'2}}\varphi_*''
 +(1-w)\varphi_*' =-\alpha(\varphi_*)(1-3w).
\end{eqnarray}
As emphasized in Ref.~\cite{dn1}, this is the equation of motion
of a point particle with a velocity dependent inertial mass,
$m(\varphi_*)=2/(3-\varphi_*^{'2})$, evolving in a potential
$\alpha(\varphi_*)(1-3w)$ and subject to a damping force,
$-(1-w)\varphi_*'$. During the cosmological evolution the field is
driven toward the minimum of the coupling function. If $\beta>0$,
it drives $\varphi_*$ toward 0, that is $\alpha\rightarrow0$, so
that the scalar-tensor theory becomes closer and closer to general
relativity. When $\beta<0$, the theory is driven way from general
relativity and is likely to be incompatible with local tests
unless $\varphi_*$ was initially arbitrarily close to 0. Thus, we
will restrict our analysis to $\beta>0$.

We need to consider three regimes~: (i) deep in the radiation era,
(ii) the effect of particle annihilation during the radiation
era (electron-positron annihilation in particular) and
(iii) the transition between the radiation and matter era.

\subsubsection{Deep radiation era}

Deep in the radiation era, $w=1/3$ and the  coupling to
$\varphi_*$ is not efficient. The equation of evolution reduces to
\begin{eqnarray}\label{kgrdu}
 \frac{2}{3-\varphi_*^{'2}}\varphi_*''
 +\frac{2}{3}\varphi_*' = 0.
\end{eqnarray}
This can be integrated to give
\begin{eqnarray}\label{kgrdusol}
 \varphi_* = \varphi_{*i} -\sqrt{3}\ln\left[
 \frac{\alpha_i\hbox{e}^{-(p-p_i)}+\sqrt{1+\alpha_i^2\hbox{e}^{-2(p-p_i)}}}
 {\alpha_i+\sqrt{1+\alpha_i^2}}
 \right]
\end{eqnarray}
where $\alpha_i$ is defined by
\begin{equation}
 \alpha_i = \frac{\varphi'_{*i}}{\sqrt{3-{\varphi'_{*i}}^2}}
\end{equation}
and where $\varphi_{*i}$ and $\varphi_{*i}'$ are the values of
$\varphi_{*i}$ and its $p$-derivative at the initial time $p_i$.
From Eq. (\ref{actionEF}), $\varphi_*$ is expressed in Planck
units and we will  allow values of $\varphi_{*i}$ to be of order
unity. Interestingly, we see that in the radiation dominated era,
$\varphi_*$ rapidly tends to a constant value. The field
derivative is just \beq
 \varphi_*' =
 \sqrt{\frac{3}{1+\alpha_i^2}}\alpha_i\hbox{e}^{-(p-p_i)}
 \eeq
so that it is divided by $e^n$ in $\Delta p=n$ e-folds. In
particular if we send $p_i\rightarrow -\infty$ then its variation
between $p_i$ and some time in the radiation era is \beq
\Delta\varphi_*\rightarrow
-\sqrt{3}\ln\left(\alpha_i+\sqrt{1+\alpha_i^2} \right). \eeq It
follows that, as long as $\varphi'_{*i}\ll\sqrt{3}$,
$|\Delta\varphi_*|\sim\varphi'_{*i}$ and the field gets frozen at
a constant value during radiation era. These properties can be
recovered easily from the form of Eq. (\ref{kgEF}) of the
evolution equation since it implies that $\dot\varphi_*$ decreases
as $\scalefac_*^{-3}$. This behavior is quite general for
dilaton-like fields \cite{condil}. In conclusion, deep in the
radiation era (much before nucleosynthesis) the initial condition
can be chosen to be $\dot\varphi_{*\xin}=0$ and
$\varphi_{*\xin}=$constant.

\subsubsection{Mass thresholds}

The previous analysis ignores an interesting effect~\cite{dpol}
that appears when the universe cools below the mass of some
species $\chi$, $T\sim m_\chi$. This species becomes
non-relativistic and induces a non-vanishing contribution to the
r.h.s. of Eq.~(\ref{kgEF}).  For example, during electron-positron
annihilation, the r.h.s. of Eq.~(\ref{kgEF})  depends on
$\Sigma_e= (\rho_e-3P_e)/\rho_{\rm rad}$.

\begin{figure}[htb]
 \center\includegraphics[width=8.5cm]{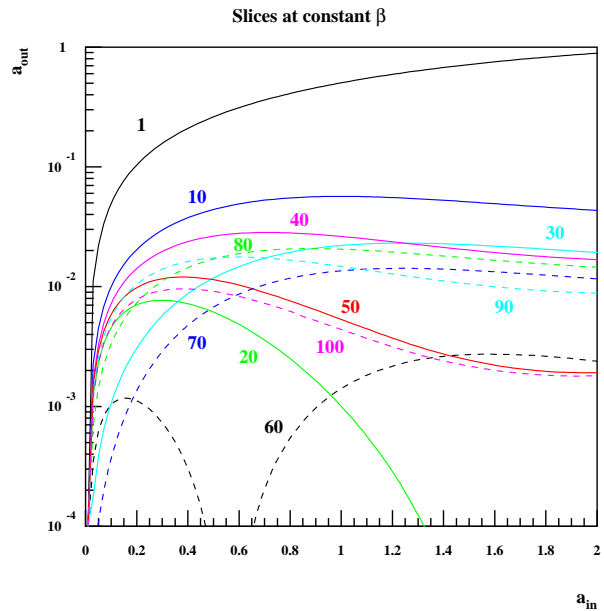}
 \caption{$a_\xout$ as a function of $a_\xin$ for different values of
 $\beta$ between 1 and 100. We see that $a_\xout<a_\xin$ which
 reflects the attraction towards general relativity during electron-positron
 annihilation.} \label{f:inout}
\end{figure}

\begin{figure}[htb]
\center\includegraphics[width=8.5cm]{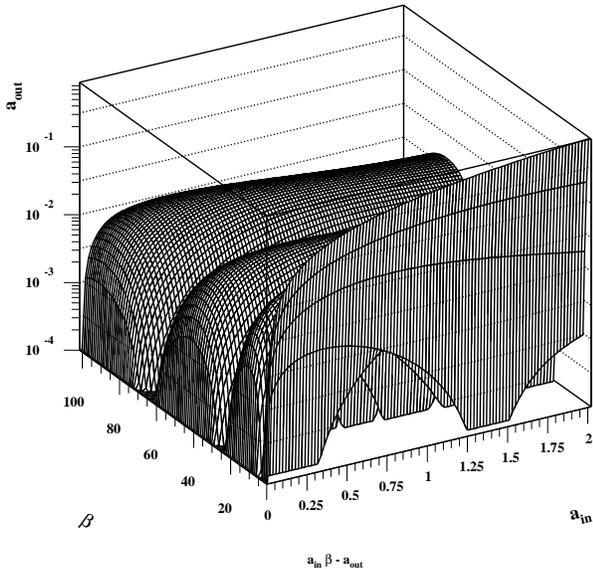}
 \caption{The general structure of
 $a_\xout$ as a function of $a_\xin$ and $\beta$.
 This illustrates the complexity of the solutions of
 Eq.~(\ref{kgqq})}.
 \label{f:inout2}
\end{figure}

In the Jordan frame the total energy density of the radiation is
\begin{equation}
\rho_{\rm rad} = g_*(T)\frac{\pi^2}{30}T^4
\end{equation}
where $g_*$ is the effective number of relativistic degrees of
freedom,
\beq
 g_*(T)=\frac{7}{8}\sum_{\rm fermion}g_i+\sum_{\rm
bosons}g_i, \eeq and $T$ is the Jordan frame temperature of the
radiation, as long as the particles are in thermal equilibrium
with the radiation bath. The term $\rho_e-3P_e$ takes the
form~\cite{cosmogen}
\begin{eqnarray}
 \rho_e-3P_e &=& \frac{g_e}{2\pi^2} m_e^2\int_0^\infty
 \frac{q^2}{\hbox{e}^{E/T}+1}\frac{\dd q}{\sqrt{q^2+m_e^2}}.
\end{eqnarray}
Intoducing $x\equiv E/T$ and $z_e\equiv m_e/T$, we deduce that
\begin{eqnarray}
 \Sigma_e(T) &=& \frac{15}{\pi^4}\frac{g_e}{g_*(T)}z_e^2\int_{z_e}^\infty
 \frac{\sqrt{x^2-z_e^2}}{\hbox{e}^x+1}\dd x
\end{eqnarray}
so that the Klein-Gordon equation~(\ref{kgqq}) takes the form
\begin{eqnarray}\label{kgqq1}
 \frac{2}{3-\varphi_*^{'2}}\varphi_*''
 +\frac{2}{3}\varphi_*' + \Sigma_e(T)\beta\varphi_*=0.
\end{eqnarray}
The force term depends on the temperature which depends on
$A(\varphi_*)$ and $p$. When this term is no longer effective, the
field evolves according to Eq.~(\ref{kgrdu}) and tends to another
constant, $\varphi_{*\xout}$. The relation between
$\varphi_{*\xin}$ and $\varphi_{*\xout}$, or equivalently between
$a_\xin=a(\varphi_{*\xin})$ and $a_\xout$, has a complicated
structure. Eq.~(\ref{kgqq}) is almost a damped oscillator (because
of the non-linear term). When $\Sigma_e(T)\beta\varphi_{*\xin}$ is small,
the field does not have the time to oscillate while $\Sigma_e$
is non-negligible, and the relation $a_\xin$-$a_\xout$ is linear.
For larger values of $\beta$ and/or $a_\xin$ one gets damped
oscillations so that $a_\xout<a_\xin$. Figure~\ref{f:inout}
depicts the relation $a_\xin$-$a_\xout$ for various values of the
parameter $\beta$ and Figure~\ref{f:inout2} illustrates the
complexity of the full solution of this equation.

\subsubsection{Details of the field dynamics near threshold}

Let us now investigate the dynamics of the attraction toward GR
during a mass threshold in more detail. In general, the
temperature is related to the integration variable $p$ by
\begin{equation}\label{Tdep}
 T[\varphi_*,p] = T_0\frac{A_0}{A(\varphi_*)}\left[\frac{q_\gamma(T_0)}{q_\gamma(T)} \right]^{1/3}
 \hbox{e}^{-p}
\end{equation}
where $q_\gamma$ is the effective number of relativistic particles
entering the definition of the entropy,  taking into account {\it
only} particles in equilibrium with the photons.

The dependence $T[\varphi_*,p]$ and the non-linear term
$\varphi_*^{'2}/3$ make Eq.~(\ref{kgqq1}) difficult to integrate.
In regimes where $\beta$ and $\varphi_{*\xin}$ are not too large
then it can safely be approximated by
\begin{eqnarray}\label{kgqapprox}
 \varphi_*''
 +\varphi_*' +
 \frac{3}{2}\Sigma_e\left(\hbox{e}^p\right)\beta\varphi_*=0.
\end{eqnarray}
This approximate equation assumes that the field is slow rolling
and that $A(\varphi_*)$ does not vary much during the transition.
It is a linear equation in $\varphi_*$ so that its solution is
proportional to $\varphi_{*\xin}$.

\begin{figure}[htb]
\vskip0.2cm
 \center{\includegraphics[width=6.cm]{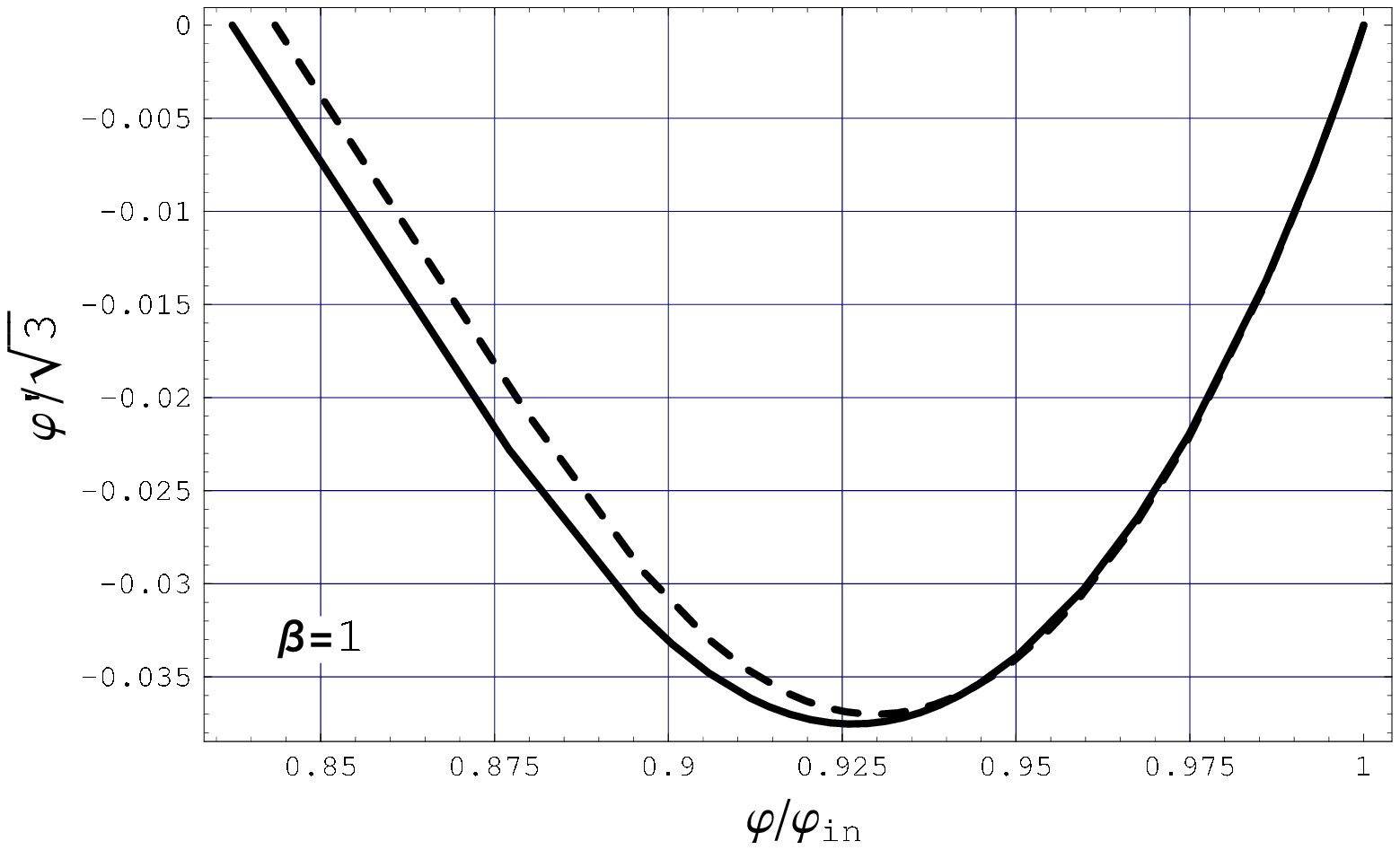}}
 \center{\includegraphics[width=6.cm]{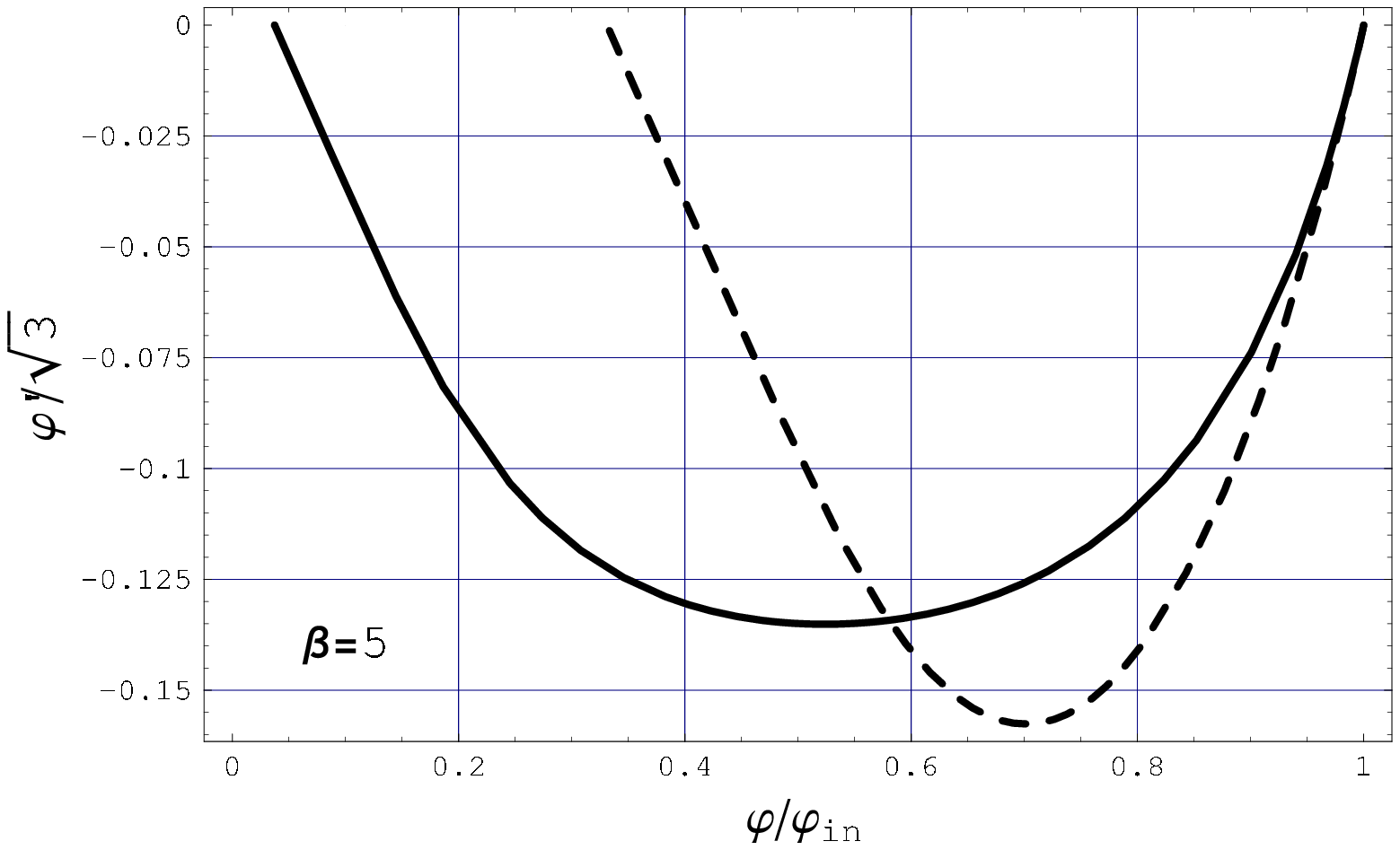}}
 \center{\includegraphics[width=6.cm]{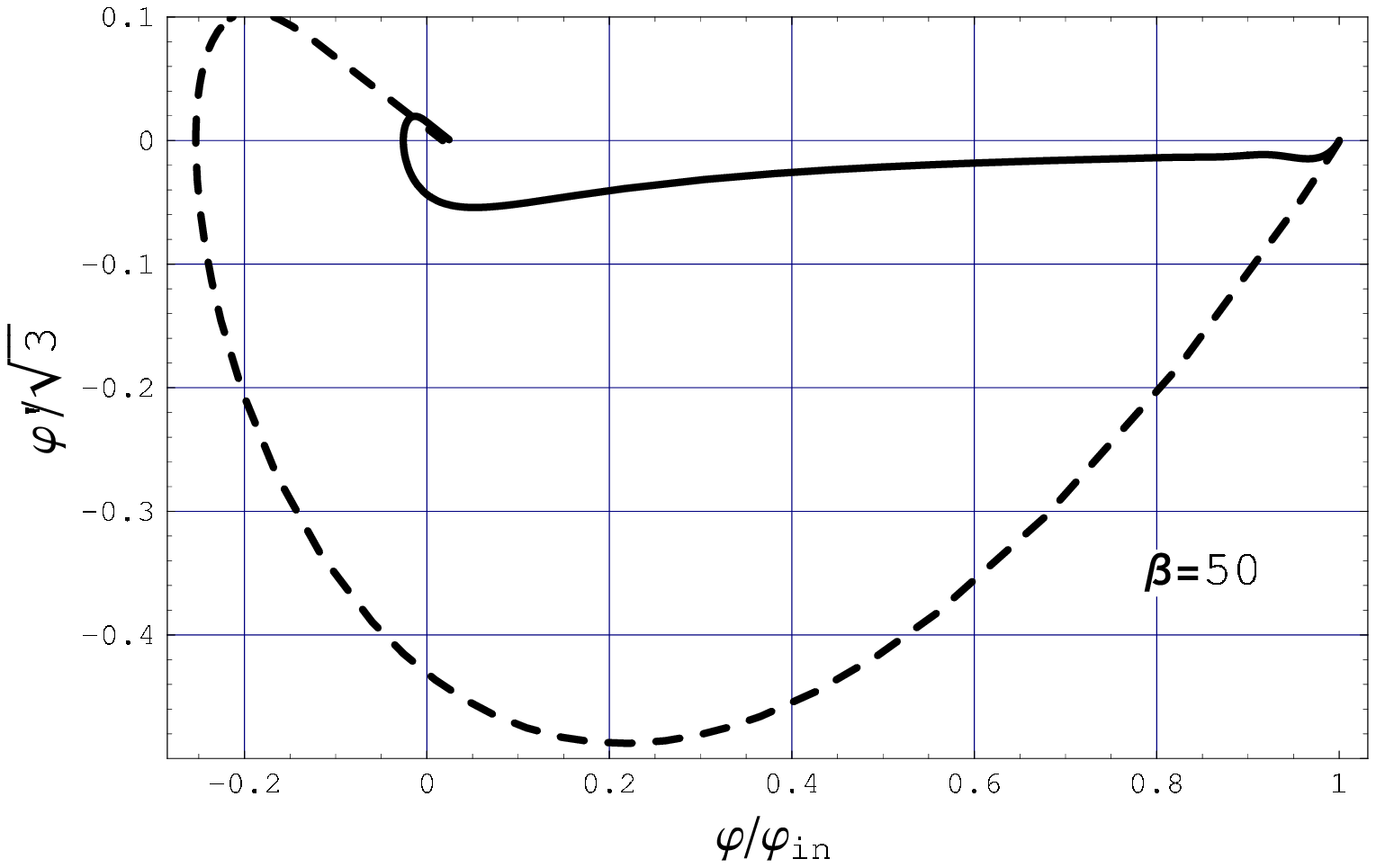}}
 \vskip0.2cm
 \caption{Evolution of the scalar field in phase space during a transition
 for $\beta=1,5,50$ from top to bottom when we assume $\varphi_{*\xin}=1$. The
 solid line corresponds to the exact solution of Eq.~(\ref{kgqq1}) while the
 dashed line corresponds to the solution of the approximate equation~(\ref{kgqapprox}).
 } \label{f:dyna1}
\end{figure}

Fig.~\ref{f:dyna1} compares the solutions of the two equations
(approximate and exact) for a single mass threshold. Indeed as
long as $\beta$ is small, the field is slow rolling and $A$
remains almost constant during the transition. This is seen in the
top panel of  Fig.~\ref{f:dyna1} for $\beta = 1$ and the field
evolves to $\varphi_* \approx 0.84 \varphi_{*\xin}$. However when
$\beta$ is large and as a consequence $A(\varphi_{*\xin})$ is also
large, the variation of $\varphi_*$ during the transition implies,
because of the relation~(\ref{Tdep}) that a given width, $\Delta
T$, corresponds to a larger $\Delta p$, while this latter is fixed
for the approximate solution. This implies that the attraction
toward $\varphi_*=0$ is more important. This progression is seen
in the middle and lower panels of Fig.~\ref{f:dyna1}.

\begin{figure}[htb]
\vskip0.2cm
 \center{\includegraphics[width=6cm]{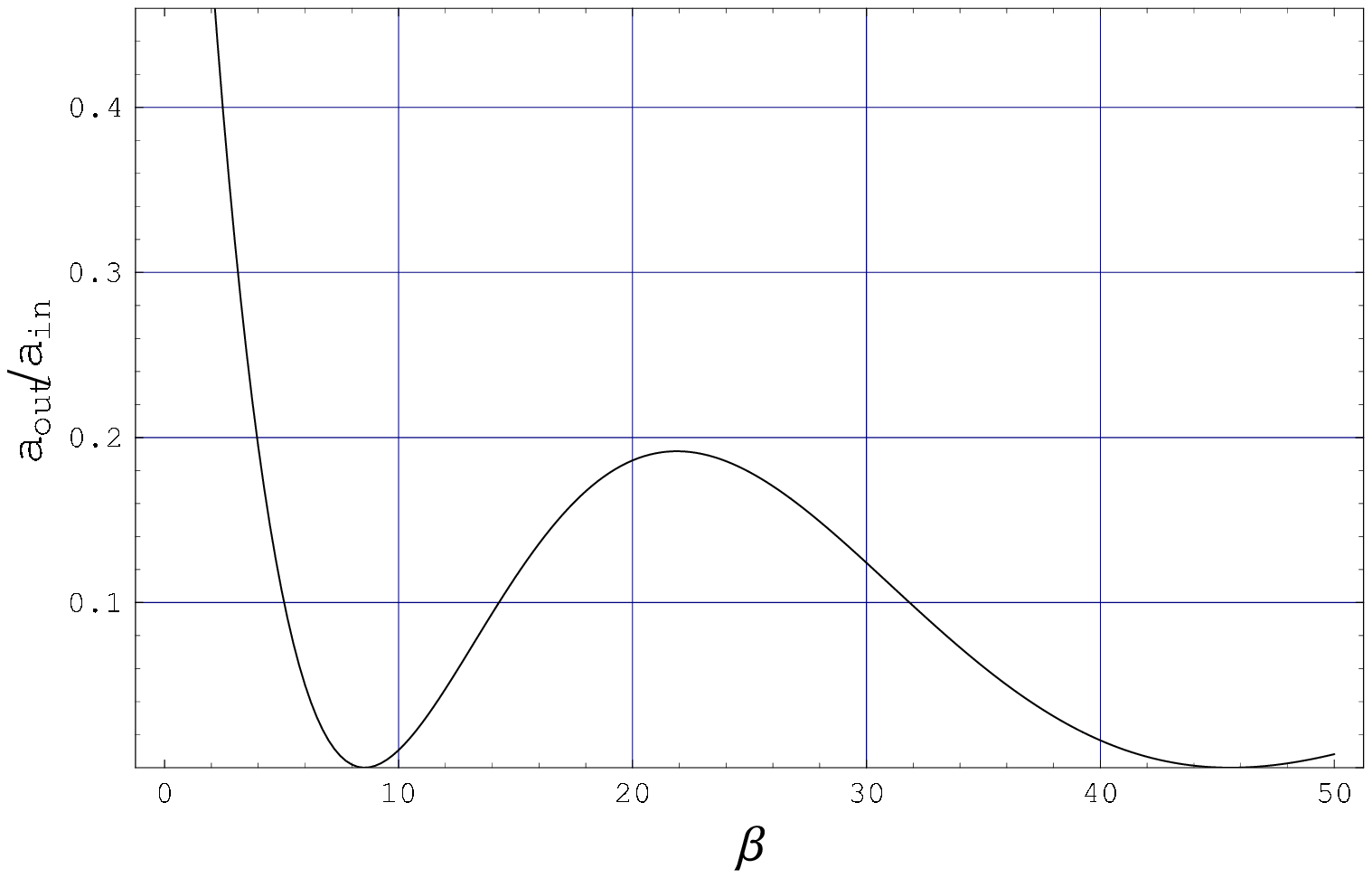}
 \center\includegraphics[width=6cm]{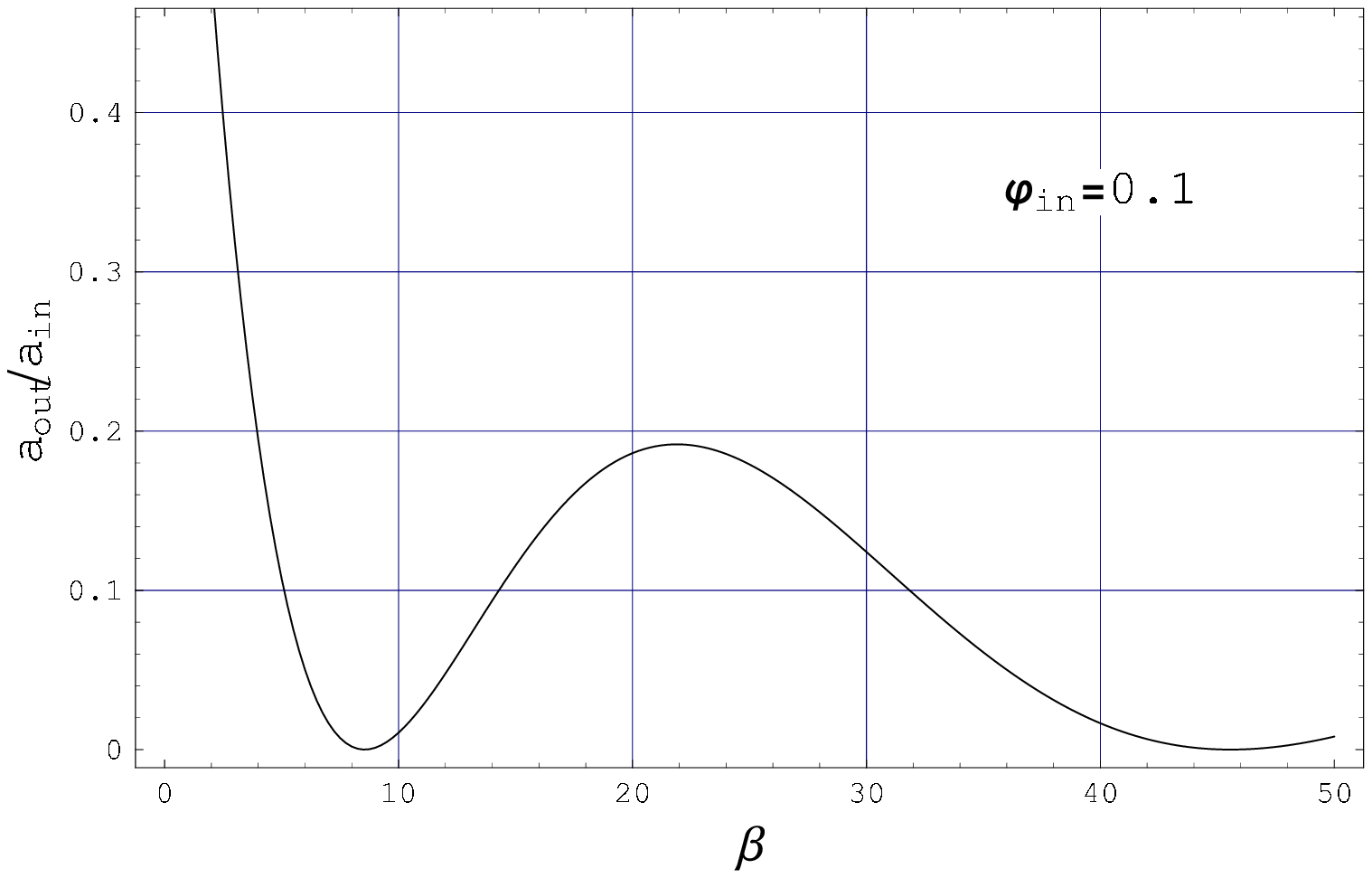}
 \center\includegraphics[width=6cm]{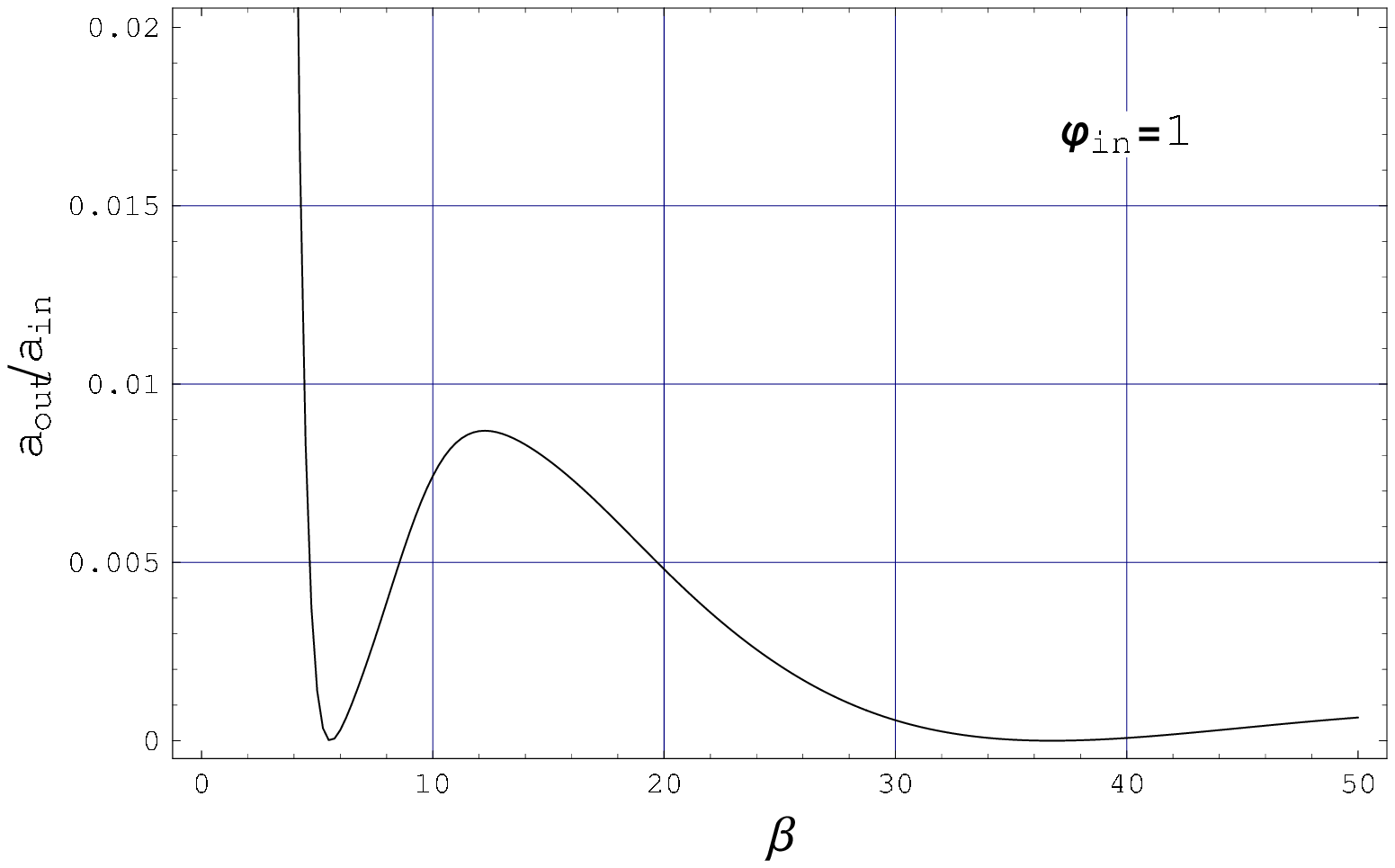}}
 \vskip0.2cm
 \caption{Evolution of $a_\xout/a_\xin$ as a function of $\beta$.
 (Top): when we used the approximate equation~(\ref{kgqapprox}),
 it does not depend on the initial value of the scalar field,
 $\varphi_{*\xin}$. (middle and bottom): we use the exact
 equation~(\ref{kgqq1}). This equation being non-linear the
 ratio depends on the initial value of $\varphi_{*\xin}$.}
\label{f:dyna2}
\end{figure}

Fig.~\ref{f:dyna2} compares the value $a_\xout/a_\xin$ as a
function of $\beta$. When Eq.~(\ref{kgqapprox}) is used, we
recover the result of Ref.~\cite{dpol}. In this case, since this
equation is linear, $a_\xout/a_\xin$ does not depend on the
initial value of $\varphi_{*\xin}$ and is a universal function of
$\beta$. This is compared to the ratio obtained from the
integration of Eq.~(\ref{kgqq1}). As long as $\varphi_{*\xin}$ is
small (typically of order 0.1), both results agree (because
$\varphi'$ remains small and $A$ does not vary significantly).
However, $a_\xout/a_\xin$ is typically 10 times smaller when
$\varphi_{*\xin}$ is of order unity. This agrees with the results
depicted in Fig.~\ref{f:inout}.

In conclusion, we see that both the depedence of the source term
for $\varphi_*$ and the non-linear term in $\varphi_*'$ lead to
significant modifications of the dynamics when the initial value of
the scalar field or $\beta$ are large.

\subsubsection{Expected value of $a_\xin$}

In the previous analysis we have restricted ourselves to
$a_\xin$ between 0 and 3, mainly for numerical reasons. We can now
in a position to justify this choice. Indeed, and as noted
earlier, because the scalar field is frozen during the radiation
dominated era, we need only specify $\varphi_{*\xin}$ as an
initial condition. For a given value of $\beta$, this fixes the
initial value of $a_\xin$.

It is difficult to predict the value of $a_\xin$ from general
arguments. For instance, if we expect $\varphi_*\sim1$ (in Planck
units) at the end of inflation, this means that
$\alpha_\xin\sim\beta$ and $a_\xin\sim\beta/2$. In this case,
one would indeed like to investigate $a_\xin$ with values up to
roughly 50. On the other hand, if we expect a deviation from
general relativity of order one at the end of inflation, then we might expect
$\alpha_\xin\sim1$, or $\varphi_*\sim\beta^{-1}$ and
$a_\xin\sim\beta^{-1}/2$. In that latter case restricting to
$a_\xin\sim3$ would be safe. Clearly,  without a detailed
model of the inflationary period it is difficult to determine the
``natural'' range of variation of $a_\xin$.

To get some insight on the expected order of magnitude of $a_\xin$
just before the period of electron-positron annihilations,  we
must investigate the effect of higher mass thresholds. To that
end, we consider an extention of Eq.~(\ref{kgqq1}) in which the
source term is replaced by a sum \beq \Sigma(T) =
\sum_{\mathrm{species}}\Sigma_i(T), \eeq where
\begin{eqnarray}
 \Sigma_i(T) &=& \frac{15}{\pi^4}\frac{g_i}{g_*(T)} z_i^2\int_{z_i}^\infty
 \frac{\sqrt{x^2-z_i^2}}{\hbox{e}^x+\epsilon}\dd x
\end{eqnarray}
with $z_i=m_i/T$, $\epsilon=+1$ for fermions and $\epsilon=-1$ for
bosons.

\begin{figure}[htb]
\vskip0.2cm
 \center\includegraphics[width=8.5cm]{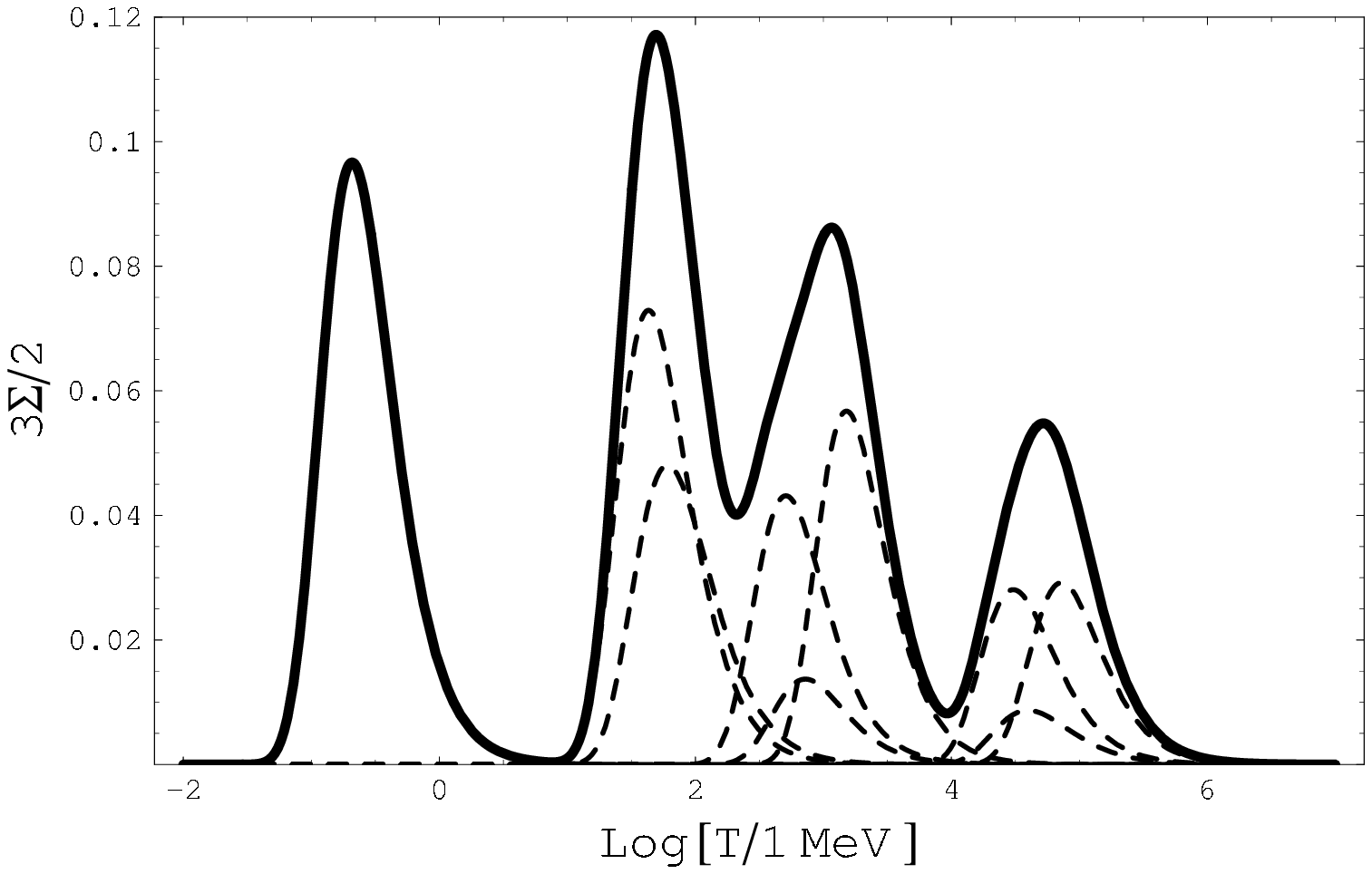}
 \vskip0.2cm
 \caption{The source function, $\Sigma(T)$, entering the Klein-Gordon equation when the
 mass thresholds corresponding to the particles listed in the text.
 The dashed curves show the individual particle contributions, $\Sigma_i(T)$, and the solid
 curve shows the sum, $\Sigma(T)$.}
 \label{f:SdeT}
\end{figure}

In principle, all massive standard model particles will play a role.
In addition to electrons and positrons, we must consider the effects of
muons, pions, charmed quarks, taus, bottom
quarks, $W^\pm$ bosons, $Z^0$ boson and the top quarks.
The role of lighter quarks is tied to the quark hadron transition
whose effect we do not include. Nor do we include the effect of the Higgs
boson due to its as yet uncertain mass.
Fig.~\ref{f:SdeT} depicts
the evolution of $\Sigma$ with (the Jordan frame) temperature. In
particular, it shows that the effects of the various thresholds
cannot be considered separately because the transitions overlap and
the scalar field, $\varphi_*$,  does not have time to settle back to $\varphi_*'=0$
between two transitions. Also note that, fortunately, the last
threshold (electron-positron annihilation) is almost decoupled
from the previous ones. Thus, we will be able to compute the state
of the scalar field just before the last transition which is of primary importance
for BBN.

\begin{figure}[htb]
 \center\includegraphics[width=6cm]{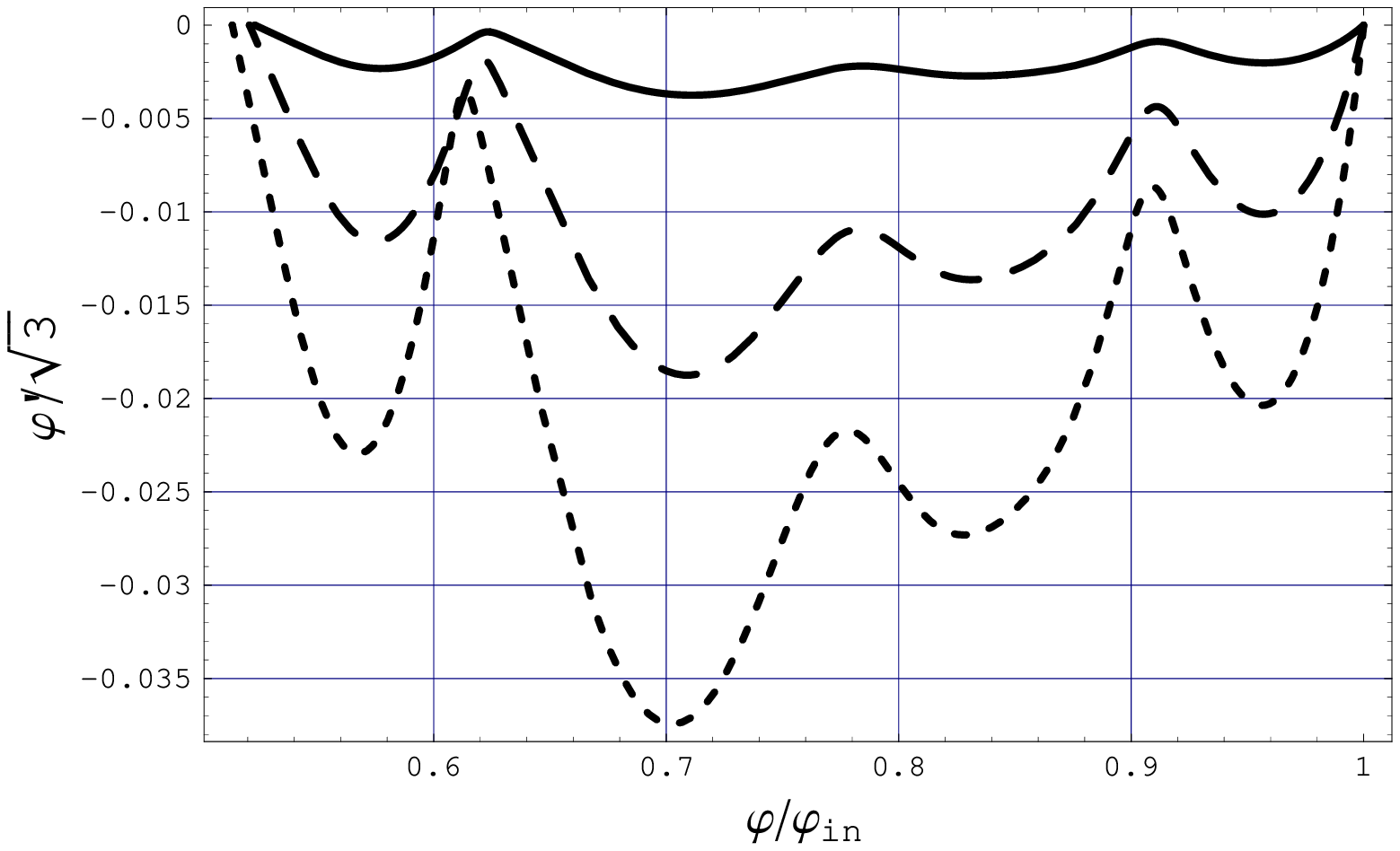}
 \center\includegraphics[width=6cm]{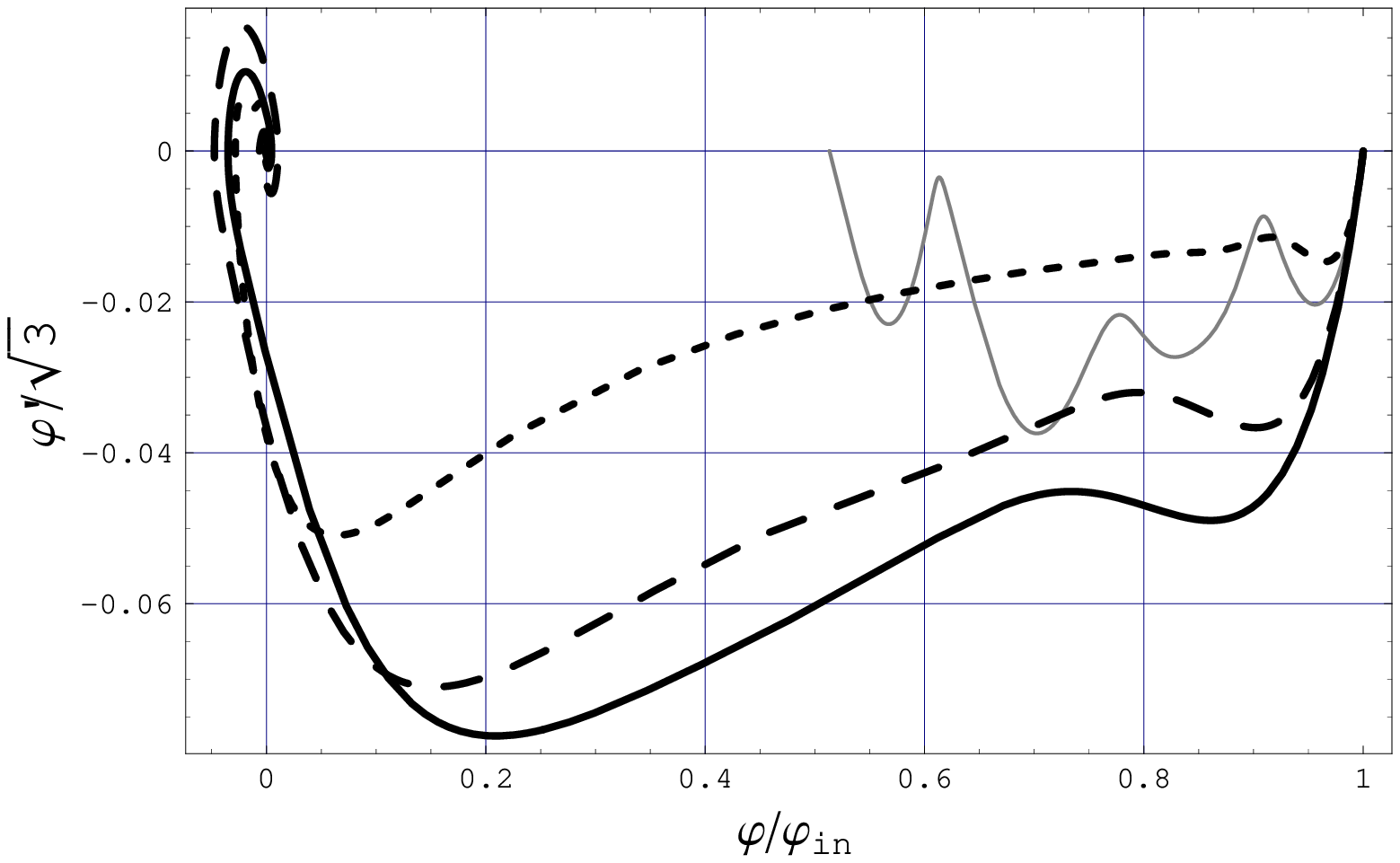}
 \caption{Dynamical evolution of $\varphi_*$ when the source term
 is described by Fig.~\ref{f:SdeT}. (Top) we have set $\beta=1$ and
 $\varphi_{*\xin}=0.1,0.5,1$ (solid, dashed, dotted).
 (Bottom) We have fixed $\varphi_{*\xin}=1$ and taken $\beta=1,15,20,50$
 (thin solid, solid, dashed, dotted).}
\label{f:dyna3}
\end{figure}

Fig.~\ref{f:dyna3} describes the dynamics of this multi-threshold
phase (including the electron-positron annihilation). As long as
$\beta$ or $\varphi_{*\xin}$ remain small, we see that each of the
four peaks of $\Sigma$, corresponds to a well defined departure
from $\varphi'_*=0$ with movement towards smaller $\varphi_*$
proportional to the initial value $\varphi_{*\xin}$. For larger
values, the field is first slow-rolling and then oscillates around
$\varphi_*=0$. In this case, we see that the effect of the four
peaks cannot be considered separately  because the field does not
have time to settle back to $\varphi_*'=0$ between two
transitions.

\begin{figure}[htb]
\vskip0.2cm
 \center\includegraphics[width=6cm]{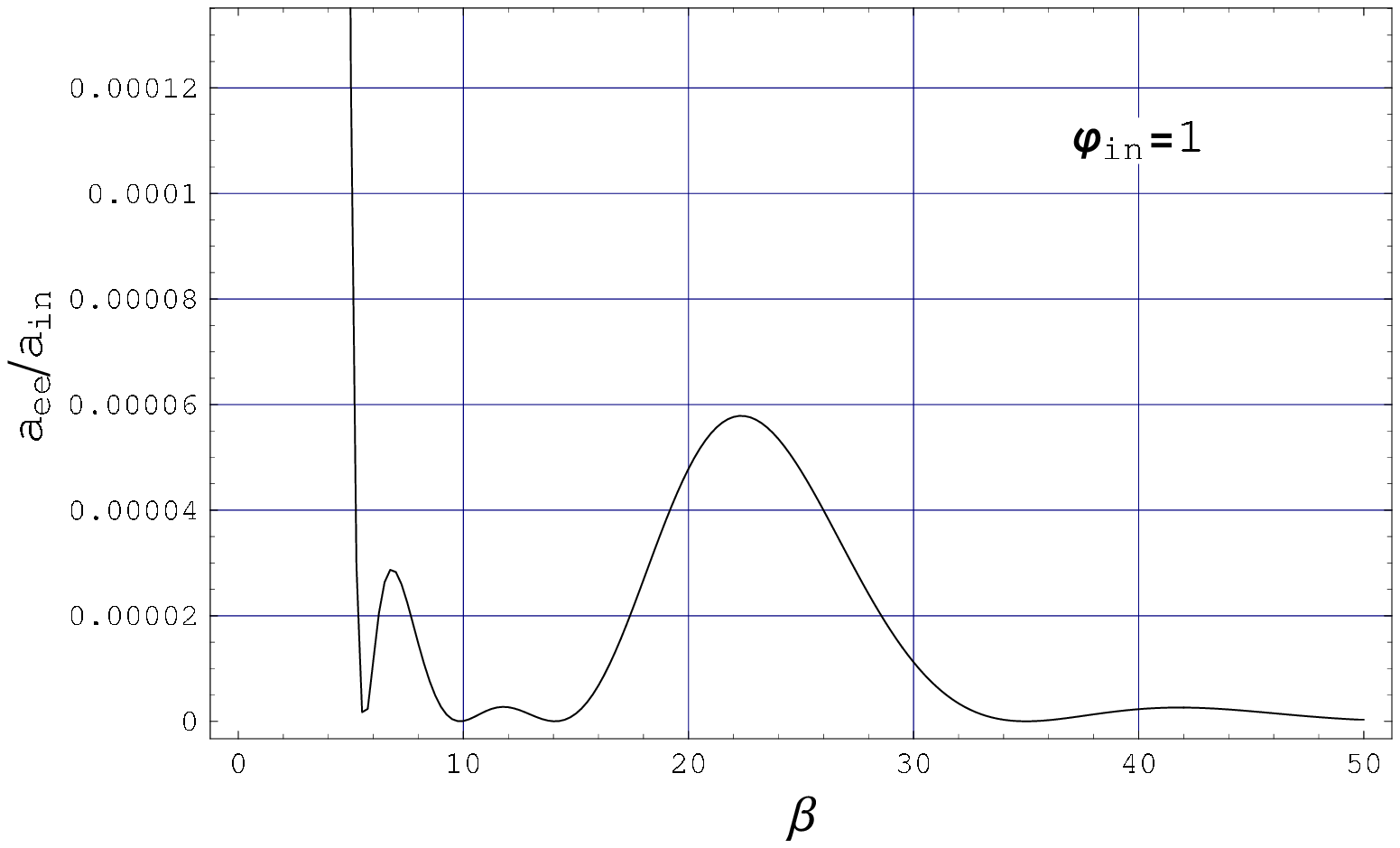}
 \center\includegraphics[width=6cm]{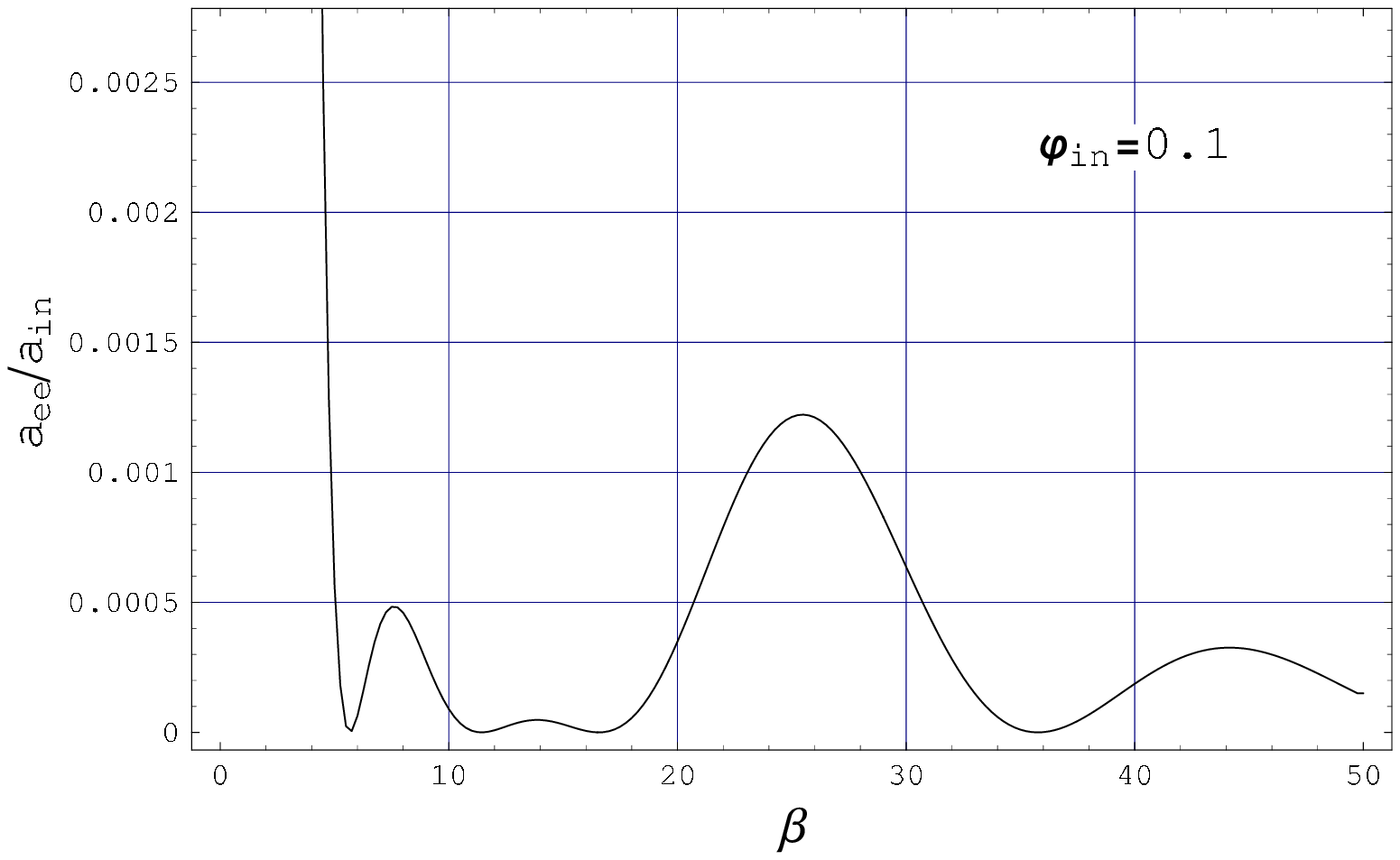}
\vskip0.2cm
 \caption{Evolution of $a_{\rm ee}/a_\xin$ as a function of $\beta$ when the source term
 is described by Fig.~\ref{f:SdeT} just before the electron-positron annihilation.
 (Top): $\beta=1$ (Bottom): $\beta=0.1$.}
\label{f:dyna4}
\end{figure}

The evolution of $\varphi_*$, when mass thresholds are
non-negligible, allows us to determine the value of $a_\xin$ prior
to the period of electron-positron annihilation.  We denote this
value by $a_{\rm ee}$. Fig.~\ref{f:dyna4} shows $a_{\rm
ee}/a_\xin$, that is the value of $a(\varphi_*)$ just before
electron-positron annihilation compared to its initial value at
very high temperature, as a function of $\beta$. The attraction
toward general relativity is very drastic. In the case where
$\varphi_{*\xin}$ is of order unity, we conclude that $a_{\rm
ee}\lesssim10^{-4}\times a_\xin\sim 10^{-4}\beta/2\lesssim
5\times10^{-3}$. It follows that restricting to $a_\xin=0 - 3$ at
before electron-annihilation is a safe limit  even if $\varphi_*
\sim {\mathcal O}(1)$ at the end of the inflationary phase.

Note also that phase transitions  are another source of attraction
toward general relativity. We have not included either the quark-hardon
transition or the electroweak transition in the
previous analysis. During a phase transition, there is
generally a significant modification of the equation of state
which will induce a source term in the
Klein-Gordon equation.

\subsubsection{Radiation-matter transition}

In principle, BBN will place a constraint on the value of $a_\xout$. As such, our
constraint will in effect be dependent on $a_\xin$ which is
unknown. To compare these constraints to the ones obtained in the
Solar system, we need to relate $a_\xout$ to $a_0$. We allow the code to
integrate the evolution equation up to the present, so that we obtain
 $a_0$ directly.

\begin{figure}[htb]
\center
\includegraphics[width=8.5cm]{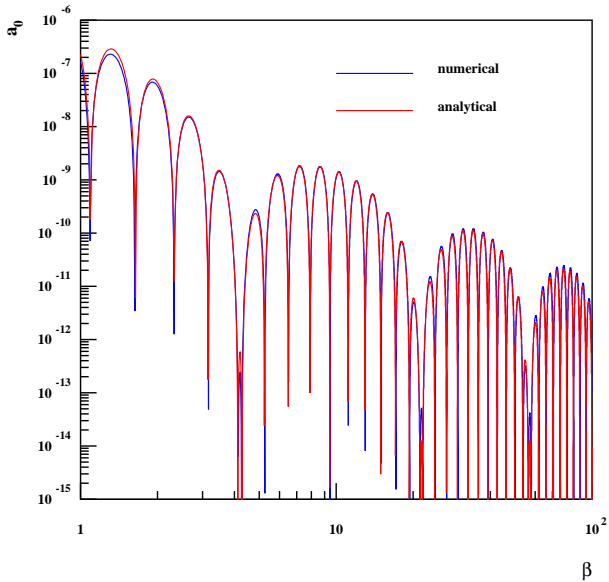}
 \caption{Comparison of the numerical integration and the analytical
 solution for a flat CDM model with $a_{\xin}=1$.}
\label{figcompare}
\end{figure}

For the particular case of a vanishing potential or as long as the field
is slow rolling, $\varphi'\ll1$, one can approximate
$3-\varphi_*^{'2}\sim3$ so that Eq.~(\ref{kgqq}) takes the
simplified form
\begin{equation}
 y(y+1)\frac{\dd^2\varphi_*}{\dd y^2} +\frac{1}{2}(5y+4)\frac{\dd\varphi_*}{\dd y}
 +\frac{3}{2}\beta\varphi_* = 0,
\end{equation}
where we have introduced the variable $y\equiv
\scalefac_*/\scalefac_{{\rm dec}*}$ and used the fact that the gas
is a mixture of presureless matter and radiation and the equation
of state is $w=1/3(1+y)$. This equation allows us to relate the
value of the scalar field deep in the radiation era but after BBN,
$\varphi_{{\rm out}*}$, to its value today, $\varphi_{0*}$. Its
solution is a hypergeometric function, $f_\beta(y)=
{}_2F_1[s,s^*,2;-y]$ with $s=3/4 -i\sqrt{3(\beta-3/8)/2}$ so that
\begin{equation}\label{e56}
  \varphi_{0*} = \varphi_{{\rm out}*} f_\beta(y_0)
\end{equation}
where the matching to the analytical solution has been performed
after the end of nucleosynthesis at a time where $\varphi_*$ is
constant. $y_0$ is given by
\begin{equation}
 y_0 = \frac{\scalefac_0}{\scalefac_{{\rm dec}}}\frac{A_{{\rm eq}}}{A_0}
     = (1+ z_{{\rm eq}})\exp[(\alpha^2_{\rm eq} - \alpha^2_0)/2\beta].
\end{equation}
This method avoids integrating the system~(\ref{s1}-\ref{s5}) to the
present but requires a determination of $y_0$. Indeed  when $\varphi$
has not varied significantly between BBN and equality, then
\begin{equation}
 y_0 \simeq (1+ z_{{\rm eq}})\exp[(\alpha^2_\xout - \alpha^2_0)/2\beta].
\end{equation}
However, this solution cannot be generalized to a $\Lambda$-CDM or
to extended quintessence models. For this reason we do not use
this method and integrate the system numerically from $z_{\rm in}$
to $z=0$. Figure~\ref{figcompare} compares our numerical
integration, from which we determine the exact value of $y_0$ and
the analytic solution~(\ref{e56}). We see that the agreement is
almost perfect. It can be checked that an error smaller than 10\%
on the evaluation of $y_0$ left $a_0$ almost unchanged. Let us
emphasize that in more general cases, i.e. for different potentials
and coupling functions, such an analytic solution is in general not
known so that the full numerical approach is necessary.

\begin{figure}[htb]
\center
\includegraphics[width=8.5cm]{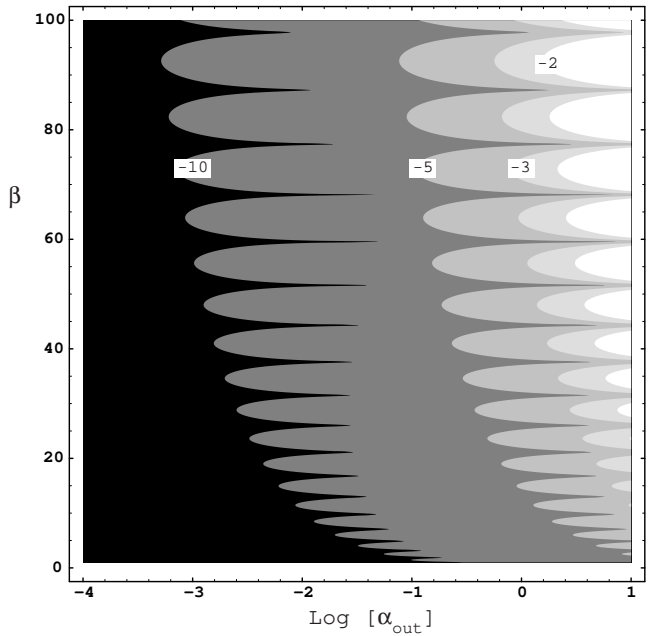}\vskip0.5cm
 \caption{$\sigma_0$ as a function of $\alpha_\xout$ and $\beta$.
 The labels on the contour lines give the value of $\log|\sigma_0|$.}
\label{figgdot}
\end{figure}

The solution~(\ref{e56}) implies that $\varphi_{*0}'=\varphi_\xout
g_\beta(y_0)$ with $g_\beta(y_0)=-3\beta y_0\,
{}_2F_1(1+s,1+s^*,3;-y_0)/4$. It is then possible to estimate,
from Eq.~(\ref{gconst0}), the value of $\sigma_0$ as a function of
($\alpha_\xout,\beta)$,
\begin{equation}
 \sigma_0 = 2\alpha_\xout g_\beta(y_0)\frac{1+\frac{\beta}{(1+\alpha_\xout^2)f^2_\beta(y_0)}}
 {1+\frac{\alpha_\xout^2}{\beta}f_\beta(y_0)g_\beta(y_0)}.
\end{equation}
As shown by Fig.~\ref{figgdot}, as soon as
$\alpha_\xout\lesssim1$, the constraint on $\sigma_0$ is
satisfied.  This means that for the quadratic coupling model, nearly all parameter choices
satisfy this constraint.

The constraint on $\sigma_0$ leading to $a_\xout < 1$ is an {\it a
posteriori} argument for not considering very large values of
$a_\xin$ at electron-positron annihilation. Very large values of
$a_\xin$ will in general lead to large values of $a_\xout$ that
are constrained by local tests on the constancy of the
gravitational constant.

\subsubsection{Equivalent speed-up factor}

\begin{figure}[htb]
\center\includegraphics[width=8.5cm]{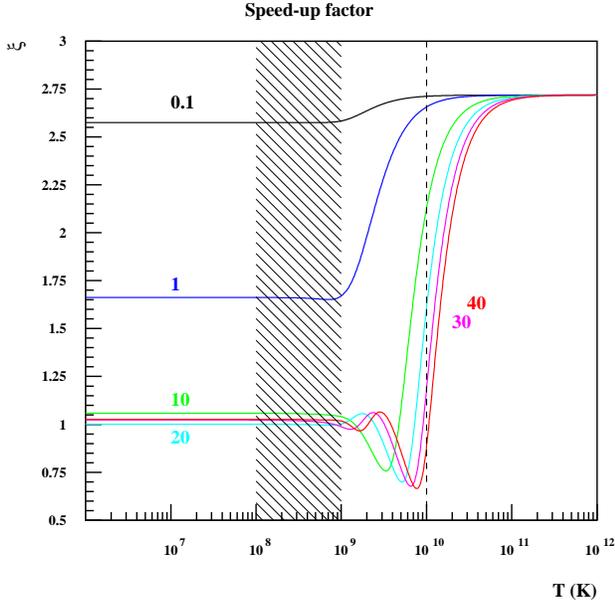}
 \caption{Variation of the speed-up factor during BBN for various values of $\beta$
 starting from the same value of $a_\xin$.
 For small values of $\beta$, $\xi$ is almost constant and is driven toward
 1 during the subsequent matter dominated era. For example, for
 $\beta\lesssim0.2$ $\xi$ is constant during BBN. For larger values of
 $\beta$, the attraction toward general relativity
 is very efficient and $\xi\sim1$ during BBN.
 The dashed line represents the time of $n/p$ freeze-out and
 we have indicated the interval during which light elements
 are formed.}
 \label{figspeed}
\end{figure}

As long as $V=0$ and one can neglect the curvature term, the
Friedmann equation~(\ref{einsteinEF1}) can be written, using
Eq.~(\ref{HHstar}) as
\begin{equation}
 3\frac{1-\varphi^{\prime2}/3}{(1+\alpha\varphi'_*)^2} H^2 = 8\pi
 G_*\rho A^2.
\end{equation}
Comparing this to the standard Friedmann equation, $3H_{GR}^2=8\pi
G_*A_0^2(1+\alpha_0^2)\rho$, one obtains the speed-up factor
 defined to be the ratio of the Hubble parameters,
\begin{equation}
 \xi =
 \frac{A(\varphi_*)}{A_0}\frac{1+\alpha(\varphi_*)\varphi'_*}{\sqrt{1-\varphi^{\prime2}/3}}
 \frac{1}{\sqrt{1+\alpha_0^2}}.
\end{equation}
Figure~\ref{figspeed} shows the variation of the speed-up
factor during BBN for various values of $\beta$, taking into
account the effects of electron-positron annihilation. $\xi$ is
constant above $z \sim 2 \times 10^{11}$ and below $z \sim 10^9$.
For large values of $\beta$, typically $\beta\gtrsim 5$, the
attraction toward general relativity is so efficient that
$\xi\sim1$ for $z \lesssim 10^9$. For smaller values, $\xi$ is
frozen at some constant value $\xi > 1$ at the end of BBN and will
be driven towards 1 only when the subsequent evolution due to
matter domination will be significant. For very small values of $\beta$, as
pointed out in Ref.~\cite{dp} and as we have shown earlier, $\varphi_*$ is almost constant
during the electron-positron annihilation period. As a result,
$\varphi_{*\xout}=\varphi_{*\xin}$ and $\varphi_*'\sim0$ so that
\begin{equation}
 2\ln\xi =\beta^{-1}(\alpha_\xin^2-\alpha^2_\xout) -
 \ln(1+\alpha_0^2).
\end{equation}
In a more complex situation, one cannot approximate this factor by
a constant and the full dynamics during BBN must be determined.
In particular, we see that $\xi$ drops around the time  the neutron-to-proton
ratio, $n/p$,
freezes out, but generically reaches a constant value during the
nucleosynthesis period.

More tuned models in which the variation of $\xi$ is not finished
during BBN may lead to some signatures on the primordial
abundances (see e.g.~\cite{larena} for a proposal). Indeed, no
model independent statements can be made but in general we expect
them to be very constrained, in particular if the mass thresholds
prior to electron-positron annihilation are taken into account.
Such models can easily be discussed in future works with the tool
presented here.

\subsection{Numerical simulations}

\begin{figure}
\center
\includegraphics[width=8.5cm]{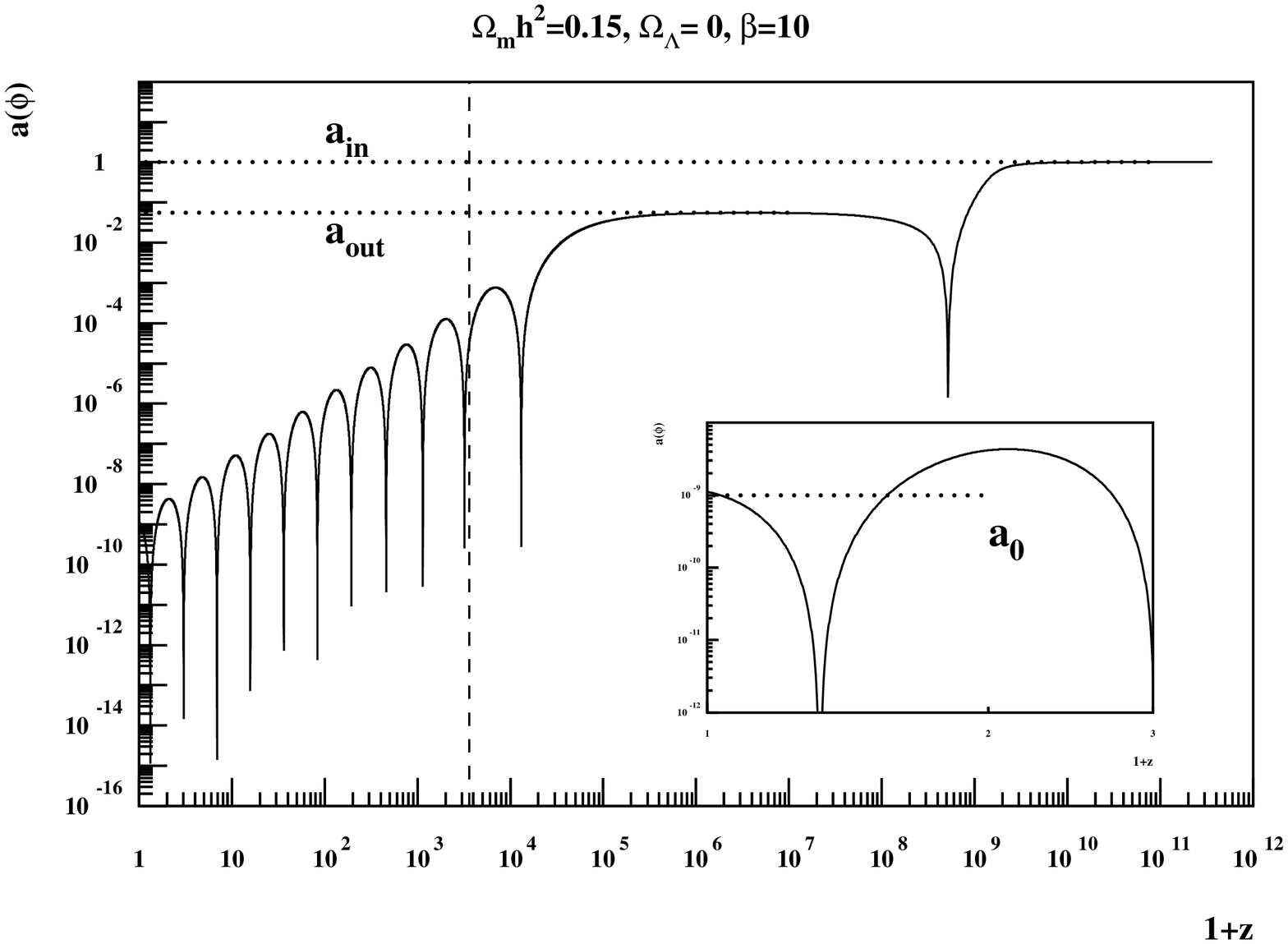}\\
\includegraphics[width=8.5cm]{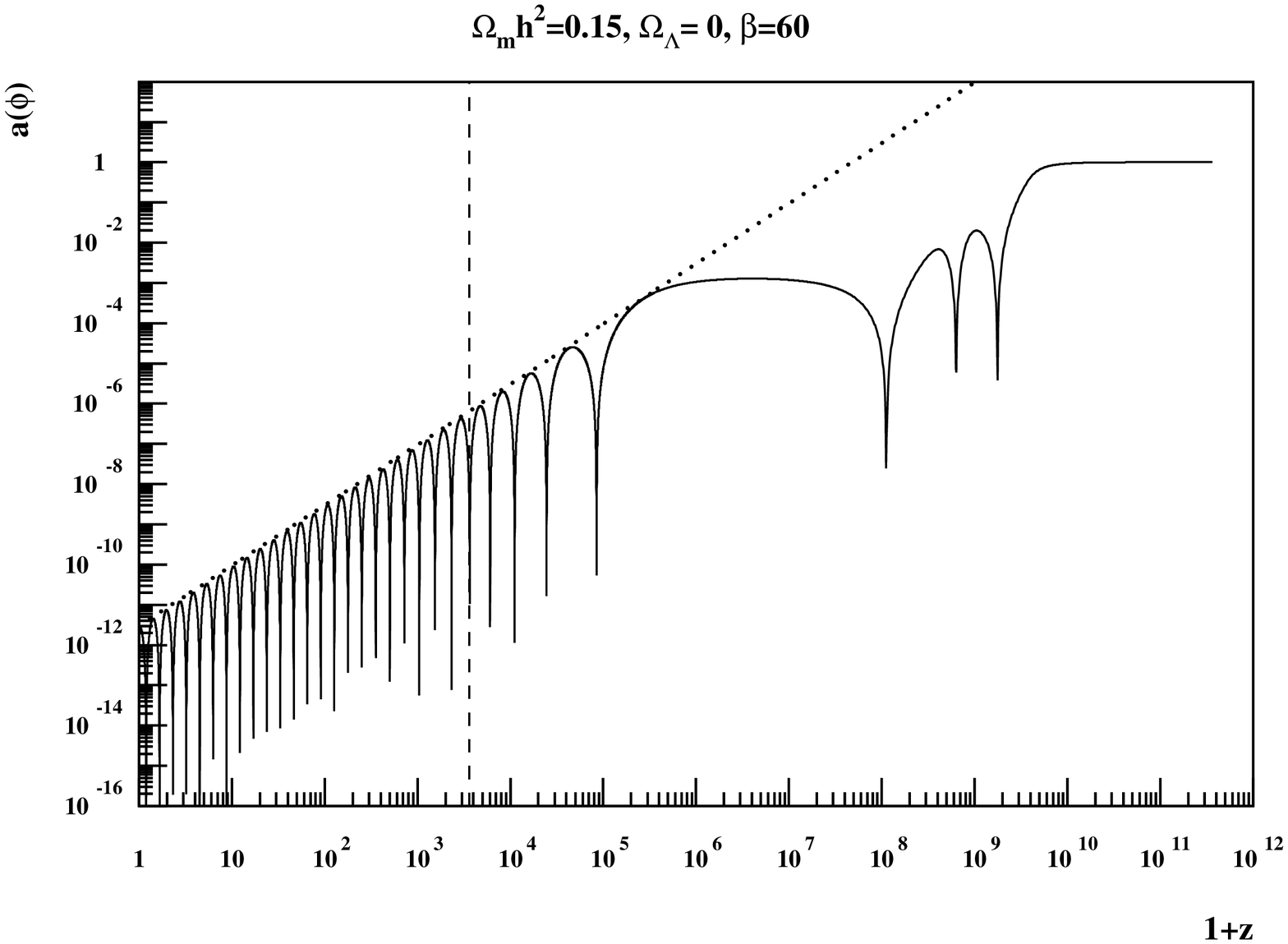}\\
\includegraphics[width=8.5cm]{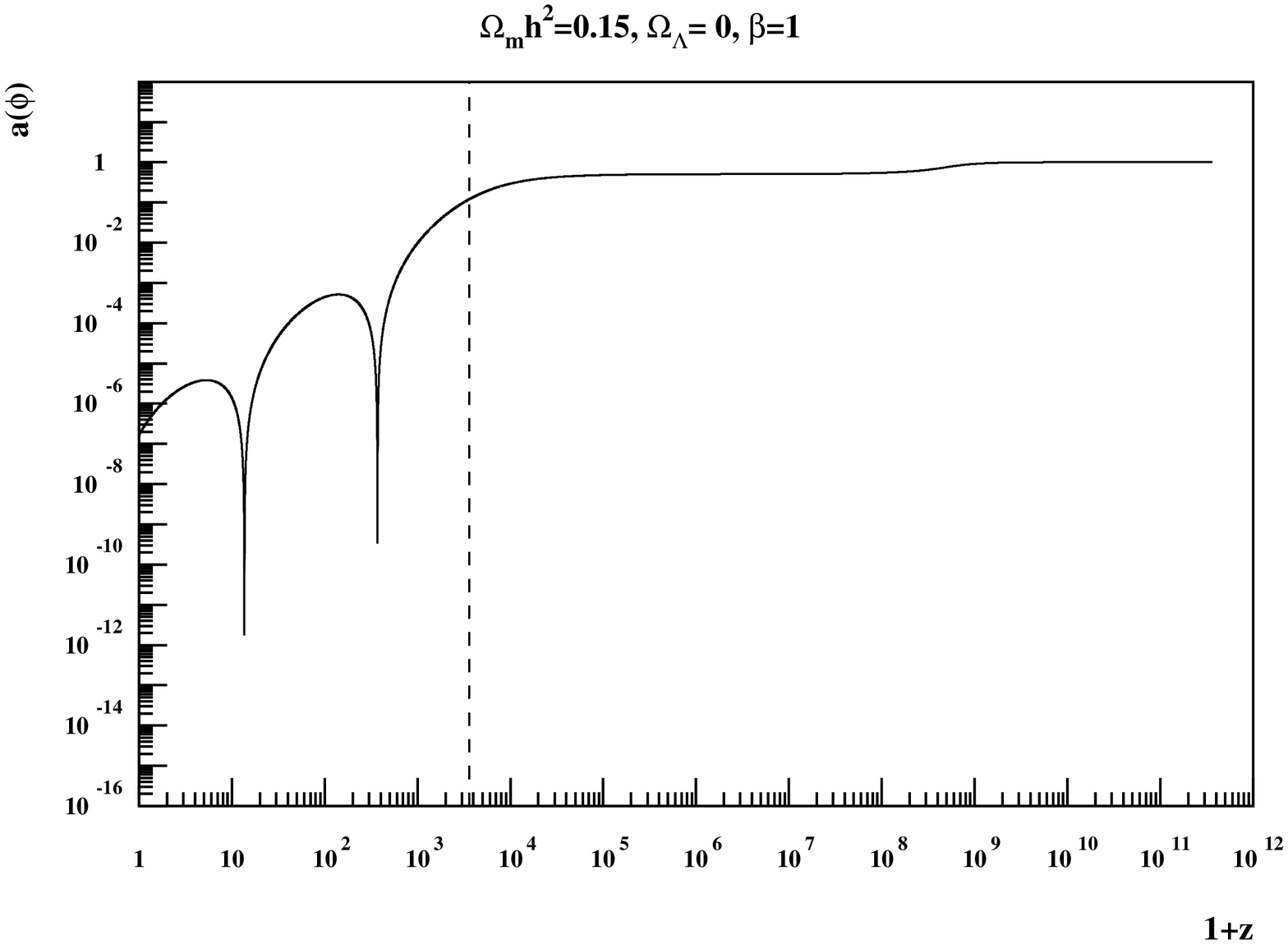}
\caption{$a(\phi_*)$ as a function of $z$ for $\beta$ = 10 (top),
$\beta=60$ (middle) and $\beta=1$ (bottom) assuming $a_{\xin}$ =
1.  The dashed lines show the redshift corresponding to
matter-radiation equality. In the top panel, the dotted lines
define $a_\xin$,   $a_\xout$ and $a_0$ while in the middle panel
it emphasizes the $(1+z)^{3\over2}$ overall dependence of
$a(\varphi)$.} \label{f:2z0}
\end{figure}

The time evolution of $a(\varphi_*)$  is depicted in
Figure~\ref{f:2z0} for three values of $\beta$. It is obtained by
numerically integrating the equations of \S~\ref{s:num} by a
standard Runge--Kutta method. In the top panel of
Figure~\ref{f:2z0}. The two plateaus at high $z$ correspond to the
constant values of $a$ during the radiation era before and after
BBN. Oscillatory behaviour due to the damped oscillation of the
field, as described by Eq.~(\ref{e56}) begins when the matter
density becomes comparable to the radiation density. In this case,
we find that  $a_\xout\approx$0.05 and $a_{0}\approx10^{-9}$. This
implies that $\alpha_0 \approx 1.4 \times 10^{-4}$.  In addition,
${\dd\varphi_*}/{\dd p}|_0 = \psi_*/H_* \approx 2 \times 10^{-5}$
leading to $\sigma_0 \simeq 7 \times 10^{-8}$, easily satisfying
the post-Newtonian constraints discussed above. For the other
examples depicted in Fig.~\ref{f:2z0}, we have
$a_\xout=1.3\times10^{-3}$, $a_{0}=2.7\times10^{-12}$, $\alpha_0
\approx -2 \times 10^{-5}$, $\psi_*/H_* \approx 4 \times 10^{-7}$,
and $\sigma_0 \simeq -9 \times 10^{-10}$ ($\beta=60$) and
$a_\xout=0.5$, $a_{0}=1.7\times10^{-7}$, $\alpha_0 \approx 6
\times 10^{-4}$, $\psi_*/H_* \approx -1 \times 10^{-3}$, and
$\sigma_0 \simeq -3 \times 10^{-6}$
 ($\beta=1$).

In the middle panel of Figure~\ref{f:2z0}, we show the evolution
of $a$ for $\beta = 60$.  As expected, the larger coupling allows
for several oscillations during $e^+ e^-$ annihilation, and
significantly more oscillations during the matter dominated era.
The dashed line shows the redshift corresponding to
matter-radiation equality. We see that the field starts
oscillating before equality due to the enhanced coupling.
Comparing the two panels, we see that  as $\beta$ increases
oscillations begin at higher redshift.

\subsection{BBN calculations}

The equations displayed in section \ref{s:num} have been
implemented in our BBN code \cite{coc,pdes}. The source term due
to the electron-positron contribution to the energy and entropy
density is calculated by a numerical integration of the fermi
distributions. The calculation starts at $T$=10$^{12}$~K well
above electron-positron annihilations and weak interaction
freeze-out with a given value of $\beta$, $a_\xin$ and
$\dd\varphi_*/\dd t_*$=0. For a given value of \obh, a grid of
calculations is performed with $a_\xin$ ranging from 0. to 3. in
steps of 10$^{-3}$ and $\beta$ ranging from 0.1 to 100. in steps
of 0.1 ($\beta<10$) and 1. ($\beta>10$). Let us emphasize that the
range in $a_\xin$ is conservative given the analysis of the
previous section. Small steps are needed because of the
complicated structure displayed in Fig.~\ref{f:inout2}. The \qua\
and \deu\ yields are compared to the allowed intervals, discussed
in the next section, and for each $\beta$ value, the maximum
allowed value of $a_\xout$ is determined and the numerical
calculation is extended to obtain the present limits on  $a_0$ and
$\alpha_0$. (The minimum allowed $a_\xout$ value was found to be
zero in all considered cases.) However, because of the late
domination of matter, the limits on $a_0$ display many more
oscillations than the limits on  $a_\xout$ as a function of
$\beta$. Hence, steps 100 times smaller  in $\beta$ were used for
the calculation of $a_0$ using interpolated values of the
relatively slowly varying $a_\xout$.

\begin{figure}[htb]
\center
\includegraphics[width=8.5cm]{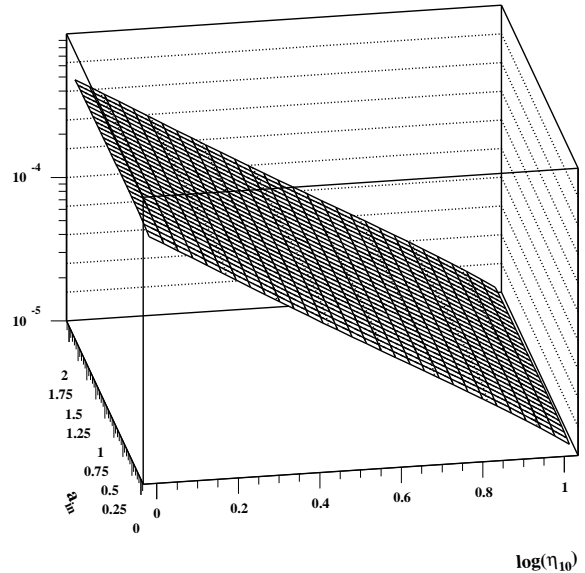}\\
\caption{\deu\ mass fraction as a function of the baryonic density
($\log(\eta_{10})$) and $a_\xin$ for $\beta$ = 20. The abundance
of D is almost insensitive to $a_\xin$.} \label{f:d3d}
\end{figure}

\begin{figure}[htb]
\center
\includegraphics[width=8.5cm]{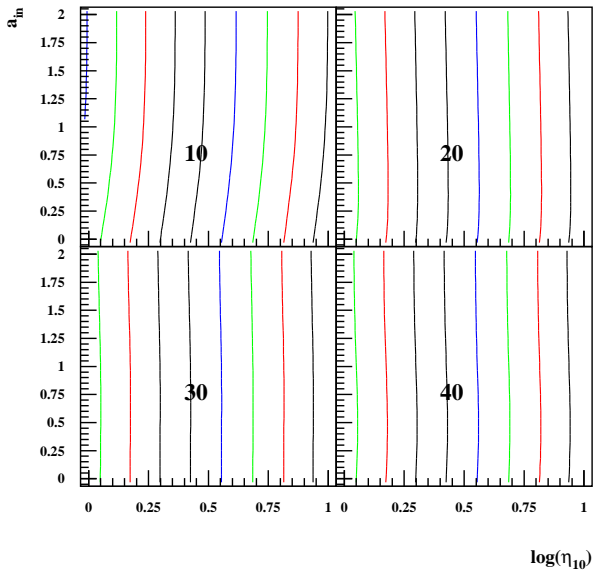}
\caption{ Contour plots for D/H as function of $a_\xin$ and
$\eta_{10}$ but for different values of $\beta$ : 10, 20, 30 and
40. (For $\beta$= 20, this is the same as
Fig.~\protect\ref{f:d3d}.) The contours are evenly spaced by step
of $\Delta\log$(D/H) = 0.2 starting from $\log$(D/H)$=-4.8$ on the
right. We conclude that the dependence of \deu\ mass fraction on
$\beta$ is mild. }
 \label{f:d2d}
\end{figure}

\begin{figure}[htb]
 \center
 \includegraphics[width=8.5cm]{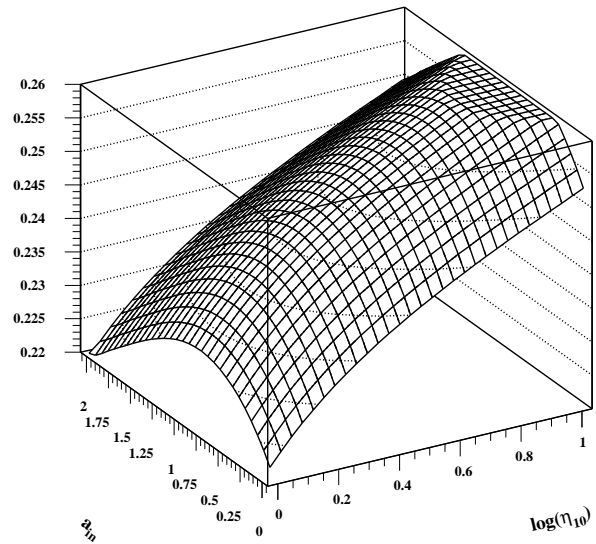}
 \caption{As in Figure~\protect\ref{f:d3d}  but for \qua. \qua\
 is the most sensitive element to the value of $a_\xin$.} \label{f:he3d}
\end{figure}

\begin{figure}[htb]
\center
\includegraphics[width=8.5cm]{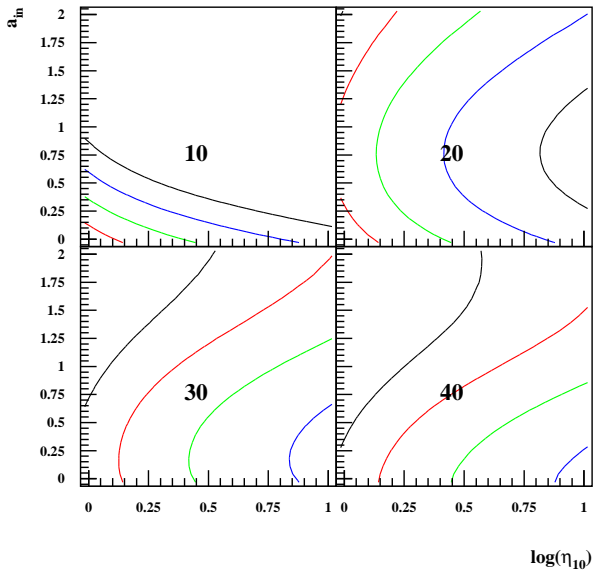}
\caption{As in Fig.~\ref{f:d2d} but for \qua. Contours correspond
from left to right to $Y_p$= 0.22, 0.23, 0.24, 0.25 and 0.26 (i.e.
black, red, green, blue and black respectively.) The abundance of
\qua\ is very sensitive to $\beta$. In addition, the dependence on
$a_\xin$ is increased for larger values of $\beta$.}
\label{f:he2d}
\end{figure}

\begin{figure}[htb]
\center
\includegraphics[width=8.5cm]{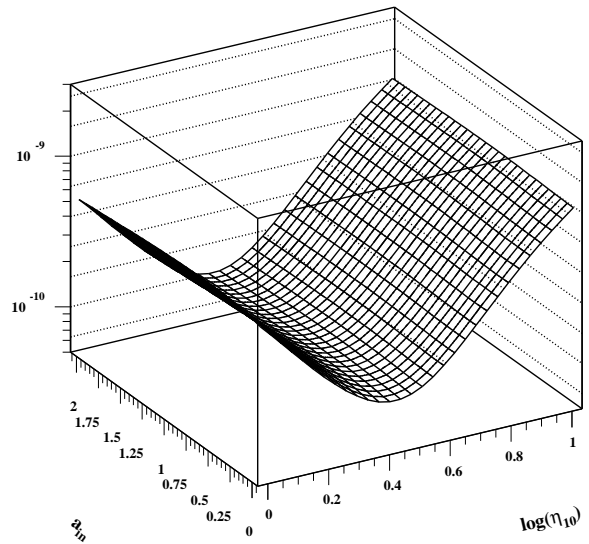}\\
\caption{As in Figure~\ref{f:d3d} but for \sep. As in the case of
deuterium, \sep\ depends very mildly on $a_\xin$.} \label{f:li3d}
\end{figure}

\begin{figure}[htb]
\center
\includegraphics[width=8.5cm]{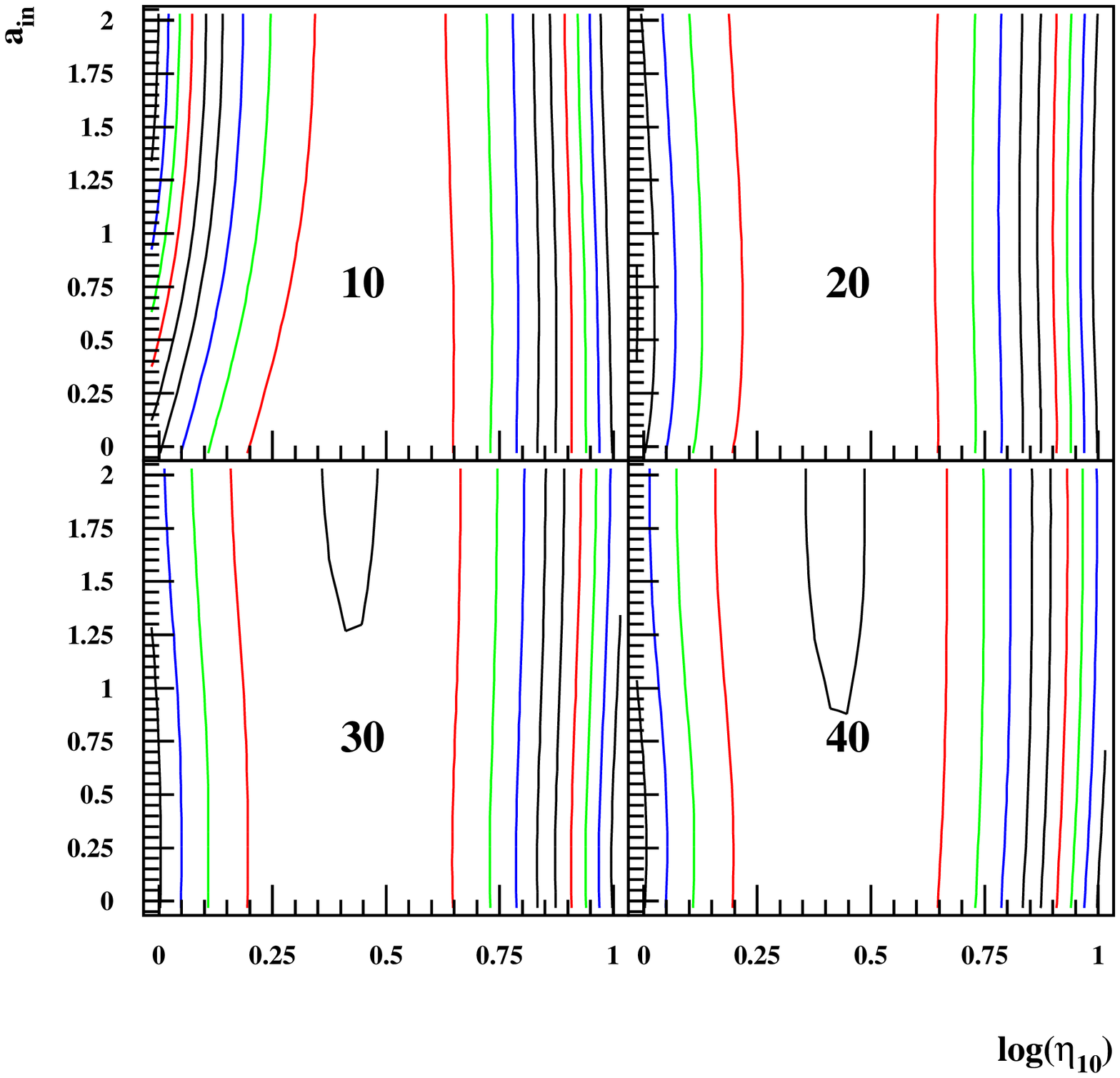}\\
\caption{As in Fig.~\ref{f:d2d} but for \sep. The contours are
evenly spaced by steps of $\Delta$Li/H=1.$10^{-10}$ starting from
the value of Li/H=1.$10^{-10}$ (in black only seen in the lower
diagrams). \sep\ shows little dependence on $\beta$ except for
values of $\log(\eta_{10})$ corresponding to the minimum value of
\sep.} \label{f:li2d}
\end{figure}

Figures~\ref{f:d3d} to~\ref{f:li2d} illustrate the dependence of
\deu, \qua\ and \sep\ in terms of the baryon-to-photon ratio,
$\eta = \eta_{10} \times 10^{-10}$, the parameter $\beta$ and the
initial condition of the field, $a_\xin$ for a model with
quadratic coupling and vanishing potential. It is evident that the
\deu\ mass fracion is only very weakly affected by the two new
parameters $(\beta,a_\xin)$. \sep\ is also almost independent of
$a_\xin$ but the valley around $\log(\eta_{10}) = 0.5$ becomes
deeper as $\beta$ increases. The most sensitive of the light
elements is \qua. Its abundance depends strongly on both
parameters. This reflects the fact that the expansion rate of the
universe is modified by this scalar-tensor theory of gravity.

\subsection{ Constraints on primordial abundances}

A comparison of the previous set of computations
with the observational determination of the light element
abundances will allow us to set constraints on this
class of models.
The abundance data are obtained from
spectroscopic observations and compared
directly with BBN predictions assuming the WMAP determination of
$\Omega_\baryon$ \cite{cfo3,coc,pdes,cuoco,cyburt}. We now discuss these observations.

\subsubsection{D/H}

The best determinations of primordial D/H are based on
high-resolution spectra in high-redshift, low-metallicity quasar absorption systems (QAS),
via its isotope-shifted Lyman-$\alpha$ absorption.
The five most precise observations of deuterium \cite{bt,omeara,kirkman,pet}
in QAS give
D/H = $(2.78 \pm 0.29) \times 10^{-5}$,
where the error is
statistical only.

Using the WMAP value for the baryon density \cite{wmap}
the primordial D/H abundance is predicted to be \cite{coc}:
\begin{equation}
({\rm D/H})_p = 2.60^{+0.19}_{-0.17} \times 10^{-5}
\label{dpred}
\end{equation}
This value is in very good agreement with the observational one.
Nevertheless, as we will see below, the agreement between
predicted D/H abundance and observations is not very sensitive to
the change the gravitational sector of the theory.

\subsubsection{\he4}

\he4 is observed  in clouds of ionized hydrogen (HII regions), the
most metal-poor of which are in dwarf galaxies. There is now a
large body of data on \he4 and CNO in these systems~\cite{iz,iz2}
for which an extended data set including 89 HII regions obtained
$Y_p$ $=$ 0.2429 $\pm$ 0.0009~\cite{iz2}.  However, the
recommended value is based on the much smaller subset of 7 HII
regions, finding $Y_p$ $=$ 0.2421 $\pm$ 0.0021.

It is important to note that \he4 abundance determinations depend on a number of
physical parameters associated with the HII region in addition to
the overall intensity of the He emission line.  These include, the
temperature, electron density, optical depth and degree of
underlying absorption. A self-consistent analysis may use multiple
\he4 emission lines to determine the He abundance, the electron
density and the optical depth. The question of systematic
uncertainties was addressed in some detail in~\cite{OSk}. It was
shown that there exist severe degeneracies inherent in the
self-consistent method, particularly when the effects of
underlying absorption are taken into account. These degeneracies
are markedly apparent when the data is analyzed using Monte-Carlo
methods which generate statistically viable representations of the
observations. When this is done, not only are the He abundances
found to be higher, but the uncertainties are also found to be
significantly larger than in a direct self-consistent approach.

Recently a careful study of the systematic uncertainties in \he4,
particularly the role of underlying absorption has been performed \cite{os2}
using a subset of the highest quality from the data of Izotov and
Thuan~\cite{iz}. All of the physical parameters listed above
including the \he4 abundance were determined self-consistently
with Monte Carlo methods. The extrapolated \he4
abundance was determined to be $Y_p = 0.2495 \pm 0.0092$ \cite{os2}.
Conservatively, it would be difficult at this time to exclude any
value of $Y_p$ inside the range 0.232 -- 0.258.

At the WMAP value for $\eta$, the \he4 abundance is predicted to
be~\cite{coc}
\begin{equation}
\label{eq:Yp}
Y_p = 0.2479 \pm 0.0004
\end{equation}
and it is in excellent agreement with the most recent analysis of
the \he4 abundance~\cite{os2}. As we will show, although  \he4
remains the most discriminatory element for physics beyond the
standard model, the current large uncertainty in its primordial
value will impede tight constraints on the parameters used to
extend minimal Einstein gravity.

\subsubsection{\li7/H}

The systems best suited for Li observations are metal-poor halo
stars in our Galaxy.  Analyses of the abundances in these stars
yields \cite{rbofn} ${\rm Li/H}|_p = (1.23^{+0.34}_{-0.16}) \times
10^{-10}$.

The \li7 abundance based on the WMAP baryon density is predicted
to be~\cite{coc}:
\begin{equation}
{\rm \li7/H} = 4.15^{+0.49}_{-0.45} \times 10^{-10}
\label{li7c}
\end{equation}

This value is in clear contradiction with most estimates of the
primordial Li abundance, as also shown by \cite{cyburt} who find :
\begin{equation}
{\rm \li7/H} = 4.26^{+0.73}_{-0.60} \times 10^{-10}
\label{li7}
\end{equation}
In both cases,  the \li7 abundance is a factor of $\sim 3$ higher than the value
observed in most halo stars.

An important source for potential systematic uncertainty stems
from the fact that the Li abundance is not directly
observed but rather, inferred from an absorption line strength and
a model stellar atmosphere. Its determination
depends on a set of physical parameters and a model-dependent analysis
of a stellar spectrum.  Among these parameters, are the metallicity
characterized by the iron abundance (though this is a small effect),
the surface gravity which for hot stars can lead to an underestimate
of up to 0.09 dex if $\log g$ is overestimated by 0.5, though this effect
is negligible in cooler stars.   The most important source for error is the
surface temperature.  Effective-temperature calibrations for stellar
atmospheres can differ by up to 150--200~K, with higher temperatures
resulting in estimated Li abundances which are higher by $\sim
0.08$~dex per 100~K.  Thus accounting for a difference of 0.5 dex
between BBN and the observations, would require a serious offset of
the stellar parameters. We note that there has been a recent analysis \cite{mr}
which does support higher temperatures, and brings
the discrepancy between theory and observations to within 2 $\sigma$.

\begin{figure}[htb]
 \vskip -2cm
 \center
 \includegraphics[width=8.5cm]{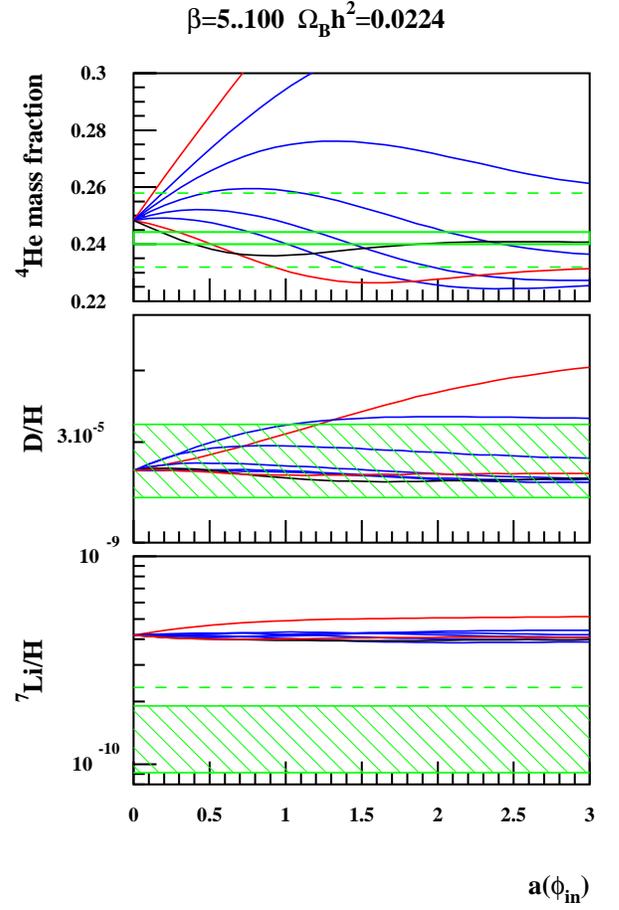}
 \caption{\qua, \deu\ and \sep\ abundances as a function of
$a_{\xin}$ for $\beta =$  5, 10, 15, 20, 25, 30, 50, and 100 (The
maximum deviations correspond to the lowest values of $\beta$).
The baryonic density is set by WMAP observations and is assumed to
be independent of the scalar component of gravity. For $\beta>10$
\deu\ is always compatible with observations and will set no
constraint while \sep\ cannot be reconciled with observations. We
conclude that BBN constraints will mainly arise from \qua.}
 \label{f:heli}
\end{figure}

We are now in a position to directly compare our numerical results
for the BBN production of light elements in a scalar-tensor theory
of gravity with observations. In Fig.~\ref{f:heli}, we show the
resulting light element abundances as a function of $a_\xin$ with
\obh=0.0224 for values of $\beta =$ between 5, 10, 15, 20, 25, 30,
50, and 100. Starting with D/H, we see from Fig.~\ref{f:heli} that
\deu\ is always compatible with observation as long as
$\beta\gtrsim10$. For lower values both \deu\ and \qua\ will set
constraints. Of course for very small beta, we must have small
values of $a_\xin$ as we approach standard GR.  In that case, the
concordance of D/H is also restored. We must also emphasize that
\sep\ cannot be reconciled with observation in this class of
models.

\begin{figure}[htb]
 \center\includegraphics[width=8.5cm]{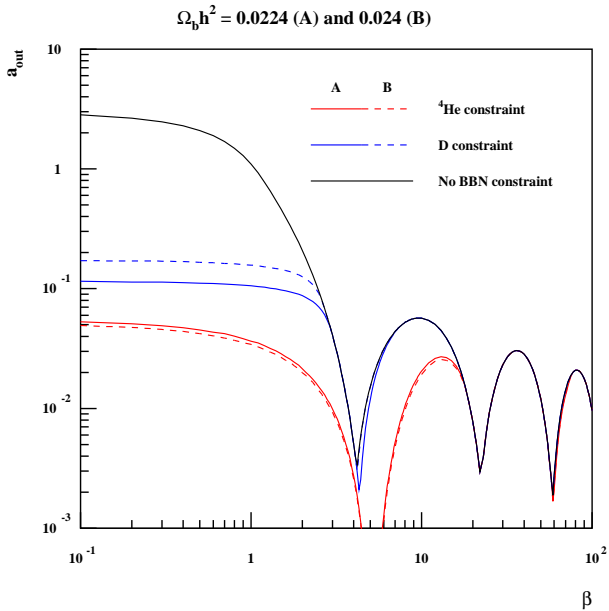}
 \caption{Constraints in the plane $(a_\xout,\beta)$ arising from
 \deu\ and \qua\ BBN abundances. The upper line corresponds to
 the maximum value reached by $a_\xout$ assuming $a_\xin=0-2$. The middle line corresponds to
 the constraint obtained when \deu\ abundances are taken into account while
 the lower line corresponds the constraint from \qua\ abundances. We conclude that
 \qua\ alone is sufficient to set the constraints on the model and that for
 $\beta\gtrsim20$, all models with reasonable $a_\xin$ are compatible
 with BBN.}
\label{f:bbnout}
\end{figure}

Fig.~\ref{f:bbnout} depicts the constraints expected on
$(a_\xout,\beta)$. The black solid curve shows the maximum
possible value of $a_\xout$ for $a_\xin = 0-2$. We also show the
maximum allowed value of $a_\xout$ from BBN for two choices of
$\Omega_b h^2$ = 0.0224 (solid) and 0.024 (dashed) based on \he4
(red) and D/H (blue). We see that for all $\beta$, \qua\ always
sets the tightest constraints. Interestingly for $\beta\gtrsim20$,
the attraction toward general relativity is so efficient that,
assuming reasonable values for $a_\xin$, all abundances are
compatible with observations.

\begin{figure}[htb]
\center\includegraphics[width=8.5cm]{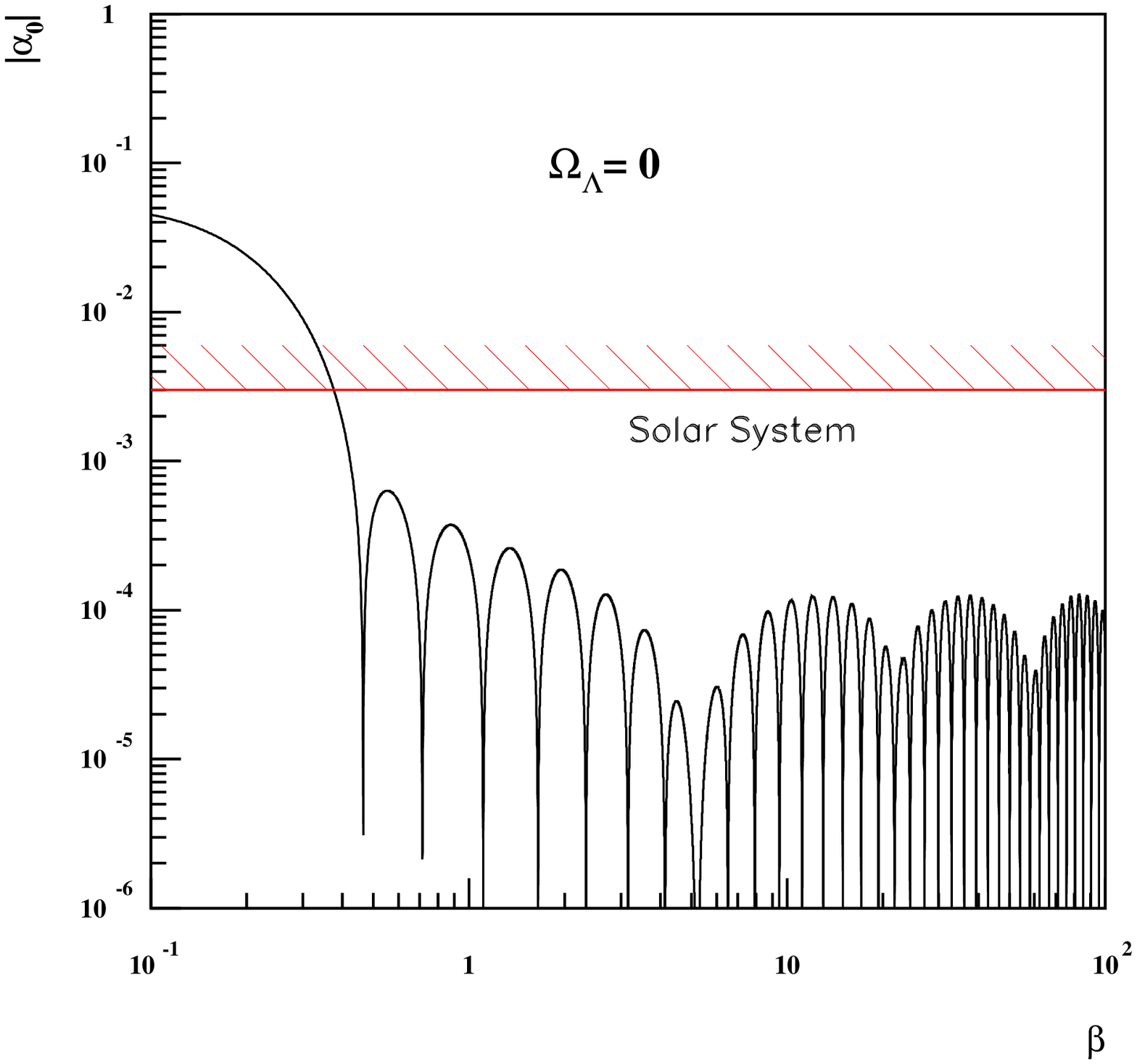}
 \caption{Constraints on ($\alpha_0,\beta)$ from \qua\ observations
 only (See Fig.~\protect\ref{f:bbnout}) using WMAP measurement
 of $\Omega_\baryon$. The dashed line represents the constraint
 obtained in the Solar System.}
 \label{f:a0n}
\end{figure}

\subsection{Using WMAP to fix $\Omega_\baryon$}

As we have seen from the previous discussion, one can set sharp
constraints on the primordial abundances if $\Omega_\baryon h^2$
is set by the analysis of the CMB anisotropies. It is important to
note however, that the WMAP data has been analyzed in a standard
cosmological set up which assumes general relativity, that is
$\alpha=\beta=0$. As was shown in Ref.~\cite{ru02}, the CMB power
spectrum in scalar-tensor theories is modified in 3 principle
ways: (1): the modification of the Friedmann equations induces a
change in the age of the universe and in the sound horizon thus
shifting the acoustic peak structure, (2) the amplitude of Silk
damping is modified because it depends on the photon diffusion
length at recombination and thus on the Hubble size at this time,
and (3) the thickness of the last scattering surface is modified.
In any specific model, one needs to check to what extent the CMB
angular power spectrum is modified and decide whether the
constraints set by WMAP can be used as is or if one needs to go
through a combined analysis to get new consistent constraints.

As was shown in Ref.~\cite{ru02}, for the case of the quadratic coupling
adopted here,
CMB anisotropies are not
affected by this modification to gravity and hence
we can safely use WMAP data to fix the baryon density. But, in general,
this will not be the case in other models (see e.g.
Ref.~\cite{carlo}).

\section{Including a cosmological constant}\label{sec_model2}

\subsection{Generalities}

The previous model does not account for the observed late acceleration of
our universe. One can easily generalize it by
introducing a cosmological constant. Let us however stress that
there is no unique way to introduce such a constant in
scalar-tensor theories.

One way to generalize the model is to introduce a cosmological constant
in the Einstein frame which corresponds to a flat
potential for the dilaton, so that the spin-0 degree of freedom
remains massless. In this case, we consider models in which
\begin{equation}
 V(\varphi_*) = V_0, \qquad a(\varphi_*) =
 \frac{1}{2}\beta\varphi_*^2.
\end{equation}
The energy density in  the Jordan frame related to the constant $V_0$ is not a
constant energy density and corresponds to a potential
$U(\varphi)=2V_0A^{-4} = 2V_0F^2(\varphi)$.

Alternatively, we can
introduce a constant energy density in  the Jordan frame. This amounts to
choosing
\begin{equation}
 V(\varphi_*) = U_0A^4(\varphi_*)/2, \qquad a(\varphi_*) =
 \frac{1}{2}\beta\varphi_*^2.
\end{equation}
The value of either $V_0$ or $U_0$ is set by the observed value of
the cosmological constant density parameter today.

The properties of these models can be discussed by generalizing
Eq.~(\ref{kgqq}) when the potential does not
vanish~\cite{bp,runaway2}. As such, we set $\rho_V=V/4\pi
G_*$ and $P_V=-\rho_V$ and $\rho_T = \rho_*+\rho_V$, $P_T = P_*
+P_V$. Using $\psi_* = H_*\varphi_*'$, we obtain
\begin{eqnarray}\label{kgqqV}
 \frac{2}{3-\varphi_*^{'2}}\varphi_*''
 +\left(1-\frac{P_T}{\rho_T}\right)\varphi_*' &=&
 -\alpha(\varphi_*)\frac{\rho_*-3P_*}{\rho_T}\nonumber\\
 &&-\alpha_V\frac{\rho_V-3P_V}{\rho_T}
\end{eqnarray}
with $\alpha_V=\dd\ln V^{1/4}/\dd\varphi_*$.

\subsection{Constant potential in the Einstein frame}

\begin{figure}[htb]
\center\includegraphics[width=8.5cm]{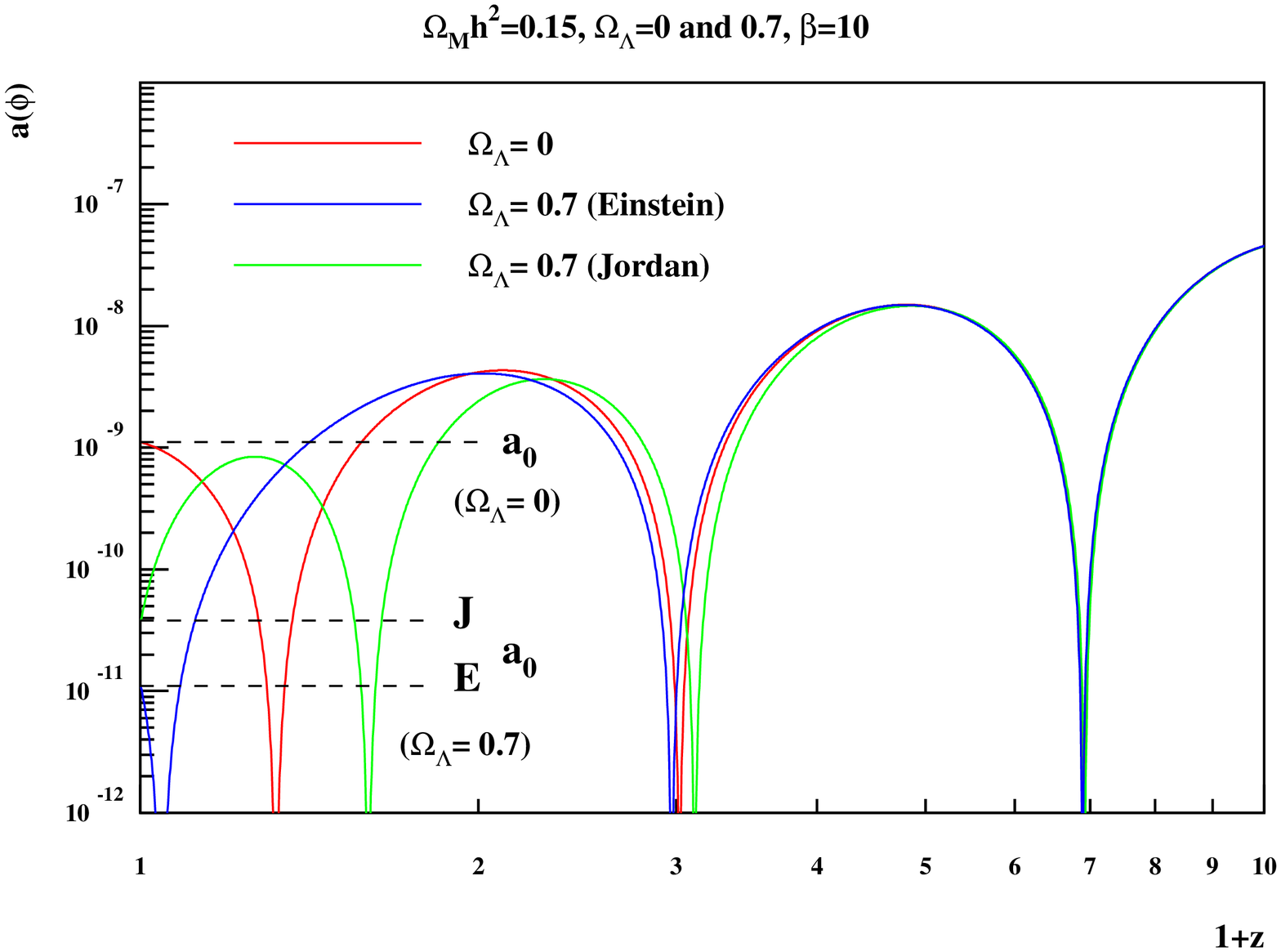}\\
\caption{Evolution of the scalar field as a function of $z$ in
 models with a vanishing cosmological constant and a cosmological
 constant defined either in Einstein or Jordan frame. We see that
 only the late time dynamics is affected by the cosmological
 constant.}
 \label{f:lambda2}
\end{figure}

In this model, the dilaton remains massless and $\alpha_V=0$. It
follows that the Klein-Gordon equation for $\varphi_*$ takes the
same form as Eq.~(\ref{kgqq}). The dynamics of the universe is
just modified at late time due to the contribution of the constant
potential to the Friedmann equation (see Fig.~\ref{f:lambda2}). It
follows that our previous results for the relation between
$a_\xin$ and $a_\xout$ are not affected and that the attraction
mechanism operates similarly. In particular the constraints on the
parameters $(a_\xout,\beta)$ obtained in Fig.~\ref{f:bbnout}
remain unchanged and the constraints on the parameters
$(\alpha_0,\beta)$ will be modified only by late time dynamics.

The value of $V_0$ is fixed by
\begin{equation}
 \Omega_{\Lambda0} = \frac{2V_0}{3H_0^2A_0^2}
\end{equation}
and it dominates only during the last $e$-fold or so. The field
is damped during the matter era so that it will be slow-rolling
close to present time. It follows that $\varphi_*''\ll\varphi'$ so that
\begin{equation}
 (1+\Omega_V) \varphi_{*0}' \sim -\alpha_0\Omega_{\rm
 mat}.
\end{equation}
Using WMAP concordance values, one finds
$\varphi_{*0}'\sim-0.2\alpha_0$ so that the
constraint~(\ref{gconst0}) on the time variation of $G_\cav$ is
satisfied, simply because $\alpha_0$ is small.

\begin{figure}[htb]
\center\includegraphics[width=8.5cm]{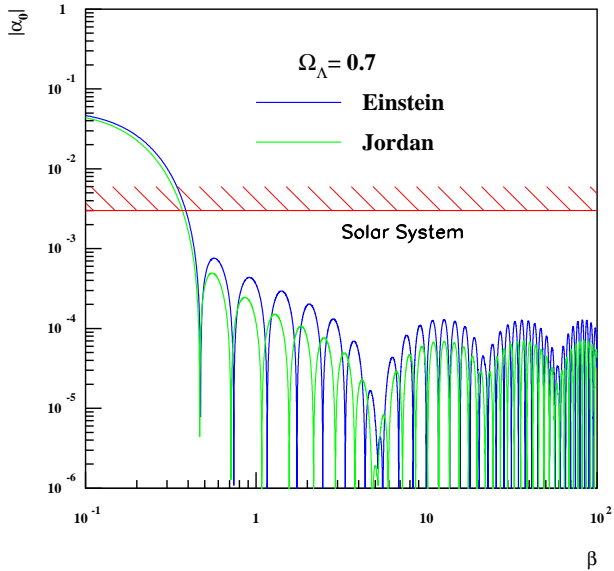}\\
 \caption{Same as Fig.~\protect\ref{f:a0n} but with a cosmological constant
 either in Einstein frame or Jordan frame.}
 \label{f:lambda1}
\end{figure}

As a consequence, we expect the effect of an Einstein-frame
cosmological constant is a shift in the global constraint contour
obtained previously when $V=0$. Figure~\ref{f:lambda1} gives the
new bounds set by BBN in the $(\alpha_0,\beta)$ plane.

\subsection{Constant potential in Jordan frame}

When one adopts a cosmological constant in the Jordan frame,
the dilaton is not massless anymore and the
value of $U_0$ is fixed by the constraint
\begin{equation}
 \Omega_{\Lambda0} = \frac{U_0A_0^2}{3H_0^2}.
\end{equation}
The coupling $\alpha_V=\alpha$. As in the previous case, the
potential will dominate only during the last $e$-fold. In the
slow-roll regime
\begin{equation}
 (1+\Omega_V) \varphi_{*0}' \sim -\alpha_0(\Omega_{\rm
 mat}+\Omega_V).
\end{equation}
With the concordance values, this implies that
$\varphi_{*0}'\sim-0.6\alpha_0$ so that, again, the
constraint~(\ref{gconst0}) on the time variation of $G_\cav$ is
satisfied. As shown in Fig.~\ref{f:lambda2}, these two models only
differ at late times and figure~\ref{f:lambda1} gives the new
bounds set by BBN in the $(\alpha_0,\beta)$ plane.

\section{Conclusions}\label{sec_concl}

This article described the implementation of general scalar-tensor
theories applicable to a BBN code. The formalism allows one to
choose any self-interaction potential as well as any coupling to
matter. As such it can be applied to models which account for the
present day acceleration such as (extended) quintessence models.

The ability to use BBN as a constraint completes our set of tools,
which include CMB anisotropies
and SNIa~\cite{ru02} and weak lensing~\cite{carlo}, to study the
cosmological imprints of this set of well-motivated theories of
gravity. All observables are computed using the same formalism for
compatibility. They can be used conjointly to set constraints on
these theories and on deviations from general relativity during
the entire evolution of our universe. We emphasize their
complementarity since BBN depends only on the background evolution
and mainly tests the attraction mechanism toward GR, CMB is mainly
sensitive to the evolution of the perturbations in the linear
regime while weak lensing probes the non-linear regime.

In this article, we have focused on the case of a quadratic
coupling in order to check our code. In the case where $V=0$ our
results are compatible with previous analysis~\cite{dp}. Note
however that they do not rely on any specific form of the analytic
solution. Also, our evaluation of the Fermi integrals that are
necessary to estimate the kick during electron-positron
annihilation do not rely on an approximation but rather on a full
numerical integration. We have used a complete BBN code
with up to date nuclear reaction rates.
Current data on the light element abundances have been used to set
constraints and we have also investigated the effect of a
cosmological constant on these constraints. For this particular model,
CMB anisotropies are not
affected and we are allowed to infer $\Omega_\baryon$ from standard CMB
analyses. We emphasize that in general this has to be checked case by case.

Since our approach is fully numerical, it can be applied to any
scenario and in particular to extended quintessence scenarios such
as models with runaway fields. In these models, during the
radiation era, the field evolves to reach a scaling solution.
Before this, there may be a kinetic phase. According to when this
kinetic phase ends, various effects on BBN can be expected. In
particular $\varphi_{*\xin}$=const. may not be a good
approximation. It was also proposed that the coupling to dark
matter may be different to the coupling to standard matter. This
hypothesis relaxes the Solar system bound and allows higher values
of $\alpha_{\rm cdm}$. All these questions, and others, will be
adressed in following works.

\noindent{\bf Acknowledgements}: We would like to thank T. Damour,
G. Esposito-Far\`ese and C. Schimd for discussions and comments.
The work of K.A.O. was supported in part by DOE grant
DE--FG02--94ER--40823. The work is also supported by the project
``INSU--CNRS/USA''.


\end{document}